\documentclass{aa}
\usepackage{psfig}

\newcommand{\HI}{\mbox{H{\sc i}\ }}
\newcommand{\HII}{\mbox{H{\sc ii}\ }}
\begin{document}
\thesaurus{01
          (11.03.01;
           11.17.4: Cloverleaf;
           12.03.3
           12.07.01
           12.12.1)}

\title{
Modelling the Cloverleaf: Contribution of a Galaxy Cluster at $z \sim 1.7$. 
\thanks{
Based on observations obtained at the Institut de Radio Astronomie
Millim\'etrique, supported by INSU/CNRS (France), MPG (Germany), 
and IGN (Spain), at the Canada France Hawaii Telescope, supported by
INSU/CNRS, the CNRC (Canada) and the University of Hawaii, and with the
NASA/ESA Hubble Space Telescope, obtained from the data archive at the
Space Telescope Institute. STScI is operated by the Association of
Universities for Research in Astronomy, Inc. under the NASA contract 
NAS 5-26555.
}
}
\author{J.-P. Kneib\inst{1}
        D. Alloin\inst{2}
        Y. Mellier\inst{3,4,1} 
        S. Guilloteau\inst{5}
        R. Barvainis\inst{6}
        R. Antonucci\inst{7}
}
\offprints{D. Alloin
           alloin@discovery.saclay.cea.fr}
\institute{
Observatoire Midi-Pyr\'en\'ees, CNRS-UMR5572, 14
Av. Edouard Belin, 31400 Toulouse, France
\and 
CNRS-URA812, Service d'Astrophysique, CE Saclay, l'Orme des Merisiers. 91191 
Gif-sur -Yvette Cedex, France
\and
Institut d'Astrophysique, 98 bis Bd. Arago, 75014 Paris, France
\and  
 Observatoire de Paris, DEMIRM. 61 Av. de l'Observatoire, 75014 Paris, France
\and
Institut de Radio Astronomie Millim\'etrique, 300 rue de la Piscine,
38406 Saint Martin d'H\`eres, France \and
MIT Haystack Observatory, Westford, MA 01886, USA \and
UC Santa Barbara, Physics Department, Santa Barbara, CA 93106, USA}
\date{Received XXX, Accepted XXX }
\maketitle
\markboth{J.-P. Kneib et al, 1997}{Modeling the Cloverleaf}

\begin{abstract} 
We present a new investigation of the Cloverleaf (z$=$2.558) based on 
the combination of archival HST/WFPC2 data, recent IRAM CO(7-6) maps 
and wide field CFHT/FOCAM images.
The deepest WFPC2 observation (F814W) shows a significant overdensity of 
I$_{814W}\sim$ 23--25 galaxies around the Cloverleaf that we interpret 
as the presence of a cluster of galaxies along the line of sight. 
The typical magnitude, red color (R-I $\sim$ 0.9) and small 
angular size of these galaxies 
suggest that the cluster is very distant and could be associated with
the absorption systems observed in the spectra of the quasar spots
(either z $=$ 1.438, 1.66, 1.87 or 2.07). The Cloverleaf is probably 
the result of the lensing effects of a system which includes, 
in addition to a single galaxy, 
one of the most distant clusters of galaxies ever detected.

With this assumption, we have modelled the lens using 
the HST/WFPC2 data and the IRAM/CO(7-6) map.
We have considered two cases: one in which the mass model is a galaxy
and a dark halo at z=1.7, and a second one in which 
the mass model is the combination of a cluster (centered on the 
overdensity of galaxies) and an individual galaxy located amid 
the Cloverleaf, both at z=1.7.
The high-resolution IRAM/CO map provides for the first time 
the orientation and the ellipticity of the CO spots induced by the 
shear component.
Velocity - positional effects are detected at the 8$\sigma$ level
in the CO map. A strong limit can then be put on the size, shape 
and location of the CO source around the quasar. The CO source is found
to form a disk- or ring-like structure orbiting the central engine at
$\sim$ 100km/s at a radial distance of $\sim$ 100pc, leading to
a central mass of $\sim$ 10$^{9}$ M$_\odot$ possibly in the form of a
black hole.
The cluster component increases significantly the convergence of the
lens and this pulls down the requirement on the mass of the lensing galaxy
by a factor 2.
This may help explain the mystery of why the lensing galaxy 
has not been detected yet. A deep high resolution infrared image
should reveal the nature and location of the lensing galaxy.\\

The presence of an additional lensing cluster along the line of 
sight to
the Cloverleaf strengthens the suspicion that many bright quasars are
 magnified by
distant clusters of galaxies at redshifts larger than 1.

\end{abstract}

\section{Introduction}

The Cloverleaf is the gravitationally lensed image of the distant quasar
H1413+117 (14$^{\rm h}$ 15$^{\rm m}$ 46$^{\rm s}$.23; 
$11^{\rm o}\ 29'\ 44''.0$ J2000.0) at
$z=2.558$ showing four spots with angular separations from $0''.77$ to
$1''.36$.  Since its discovery (Magain et al. 1988), the Cloverleaf has
been observed spectroscopically and imaged with ground based telescopes
in various bands from B to I
(Kayser et al 1990, hereafter K90, Angonin et al 1990, Arnould et al
1993) as well as at 3.6 cm with the VLA (K90). Models of the gravitational 
lens, involving one or two galaxies, were derived from these data sets, with
some emphasis put on the VLA radio map (K90). 
Recently, Keeton, Kochanek and Seljak (1997) (hereafter KKS)
 in a theoretically motivated
paper studying "shear" in gravitational lens quads system, presented
Singular Isothermal Sphere (SIS)  plus external shear lens models for most 
quads and in particular for the Cloverleaf.
The main difficulty with these models lies in the fact that 
they predict, for a z=1.44 lens, 
a mass of $\sim$ 2.5 $10^{11}$ $h^{-1}_{50}$ M$_\odot$  
within  $0.7''$ (6 $h^{-1}_{50}$ kpc) radius, which would correspond to
a relatively bright normal galaxy: 
so far, searches in the K band of the predicted `bright' 
lensing galaxies have been unsuccessful (Lawrence 1996) and this fact
remains a mystery.

HST observations of the Cloverleaf have been done in different modes:
Falco (1993) presented a short report on the pre-COSTAR  HST/WFPC image
(used for the lens model  derived by Yun et al (1997)), 
Turnshek (1996) discussed FOS observations,
and a recent preprint by Turnshek et al (1997) described
PC observations (pre- and post-COSTAR).
However, the post-COSTAR HST/WFPC2 (PC and WF) data have not been
fully exploited yet, though they may bring new constraints on the
gravitational lens modeling.  

New information also comes from the detection of the
Cloverleaf in the molecular CO transitions (Barvainis et al 1994, Wilner
et al 1995, Barvainis et al 1997, Yun et al 1997, Alloin et al 1997).
Both the physical conditions of the molecular gas in the quasar can be
studied in detail (Barvainis et al 1997) and the gravitational lens
model could be improved. Indeed, unveiling the gravitational lens nature
of the Cloverleaf in CO (Alloin et al 1997) provides the true
intensity ratios of the four spots, 
unaffected by absorption from intervening material, and
unaffected by microlensing effects as the CO source in the quasar 
is not point-like (a gravitational shear is measured on the 3
brightest CO spots A, B \& C). However, the different parts of the CO 
source suffering different amplification, the CO intensity ratios 
cannot be compared directly to the optical ones: this explains the
corresponding 
observed differences reported by Alloin et al (1997). An improved CO(7-6)
map, presented in this paper, has been obtained with the 
IRAM interferometer, giving further
insight into the geometry of the quasar CO source.

In addition, CCD images of the Cloverleaf over a 5' field of view
(CFHT/FOCAM archives) allow  precise astrometry of the field
and a better registration of the maps obtained through various wavebands,
with a  particular interest in the optical and the millimeter ones.

In the current paper, we take into account all these new constraints in
order to derive an improved model of the gravitational lens  and
of the quasar molecular source. We present in Sect. 2 the new CO(7-6)
data set and discuss in Sect. 3 the post-COSTAR HST/WFPC2 data, 
from the point of view of the 
Cloverleaf itself and of its environment. 
We provide in Sect. 4 the results of the
astrometry performed from archived CFHT/FOCAM images and, using the 
CO(7-6) high resolution map obtained with the IRAM interferometer, we 
position very accurately the optical and
millimeter data sets. A new gravitational lens model is computed and 
presented in Sect. 5, while the CO source in the quasar is discussed in
Sect. 6. 
A final discussion and prospective analysis is given in Sect. 7.
Throughout the paper, we use H$_0$= 50 $h_{50}$ km/s/Mpc, $\Omega_0$=1 and
$\Lambda=0$.

\section{Improved IRAM Interferometric CO(7-6) map}

\begin{figure*}
\caption{Image of the Cloverleaf obtained with the IRAM telescope
at Plateau de Bure.
(a) is the total CLEANed image, (c) the CLEANed blue-shifted 
image, (d) the CLEANed red-shifted
image and (b) the difference between the CLEANed red and CLEANed blue image.
The CLEANed CO(7-6) maps were obtained with a natural beam of
$0.8''\times 0.4''$ at P.A. 15 deg. They have been restored with a circular
$0.5''$ beam for comparison with HST data. 
Contour spacing is 1.35 mJy/beam for 1a, 2 mJy/beam for 1c and 1d
and 3 mJy/beam for 1b, corresponding in each case to 2$\sigma$.
}
\label{fig:iram}
\end{figure*}

\begin{table*}[t]
\begin{tabular}{lcccccc}
\hline
Spot & $\Delta \alpha$ ($''$) & $\Delta \delta$  ($''$)
& Flux (mJy) & Major axis ($''$) & Minor axis ($''$) & P.A. \\
\hline
\hline
\multicolumn{7}{c}{Velocity Range -225, +225 km/s}\\
\hline
A & 0 & 0 & 12.9  (2.1) & 0.50 (0.17) & 0.30 (0.14) & 0    (23) \\
B & 0.71 (0.026) & 0.15 (0.033) & 26.7  (1.9) & 0.60 (0.08) & 0.25
(0.12) & 70    (12) \\
C & -0.38 (0.018) & 0.63 (0.023) & 24.0  (2.0) & 0.66 (0.10) & 0.40
(0.08) & -3    (8) \\
D & 0.42 (0.025) & 1.02 (0.030) & 17.0  (1.9) & 0.56 (0.11) & 0.25
(0.20) & 73    (15) \\
\hline
\hline
\multicolumn{7}{c}{Velocity Range -225, -25 km/s}\\
\hline
A & 0 & 0 & 13.5  (2.2) & 0.49 (0.24) & 0.30 & 0 \\
B & 0.71 & 0.15 & 24.4  (2.2) & 0.48 (0.10) & 0.25 & 70 \\
C & -0.33 (0.02) & 0.56 (0.03) & 30.8  (2.1) & 0.60 (0.11) &   0.40  & 
0  \\
D &  0.42 &  1.02 &    14.9  (2.3) &  0.30 (0.16)
 &  0.25  &  75 \\
\hline
\hline
\multicolumn{7}{c}{Velocity Range +25, +225 km/s}\\
\hline
A & 0 & 0 & 10.8  (2.2) & 0.32 (0.33) &   0.30  &       0  \\
B &  0.71  & 0.15  &  27.4  (2.2) & 0.63 (0.08) &  0.25 & 70  \\
C & -0.41 (0.02) & 0.69 (0.03) &  19.1  (2.1) &  0.52 (0.16)  &  0.40  & 
33 (25) \\
D & -0.29 &  0.87 &  18.3  (2.3) &   0.60 (0.15) &  0.25 &  75 \\
\hline
\end{tabular}
\caption{The Cloverleaf CO(7-6) image: positional shifts, fluxes and
elongations of the spots. One $\sigma$ error bars are given within brackets.
When no error bar is given, the value was fixed from the fit to the 
-225,225 km/s image.
}
\label{tab:spotiram}
\end{table*}

\begin{figure}
\psfig{figure={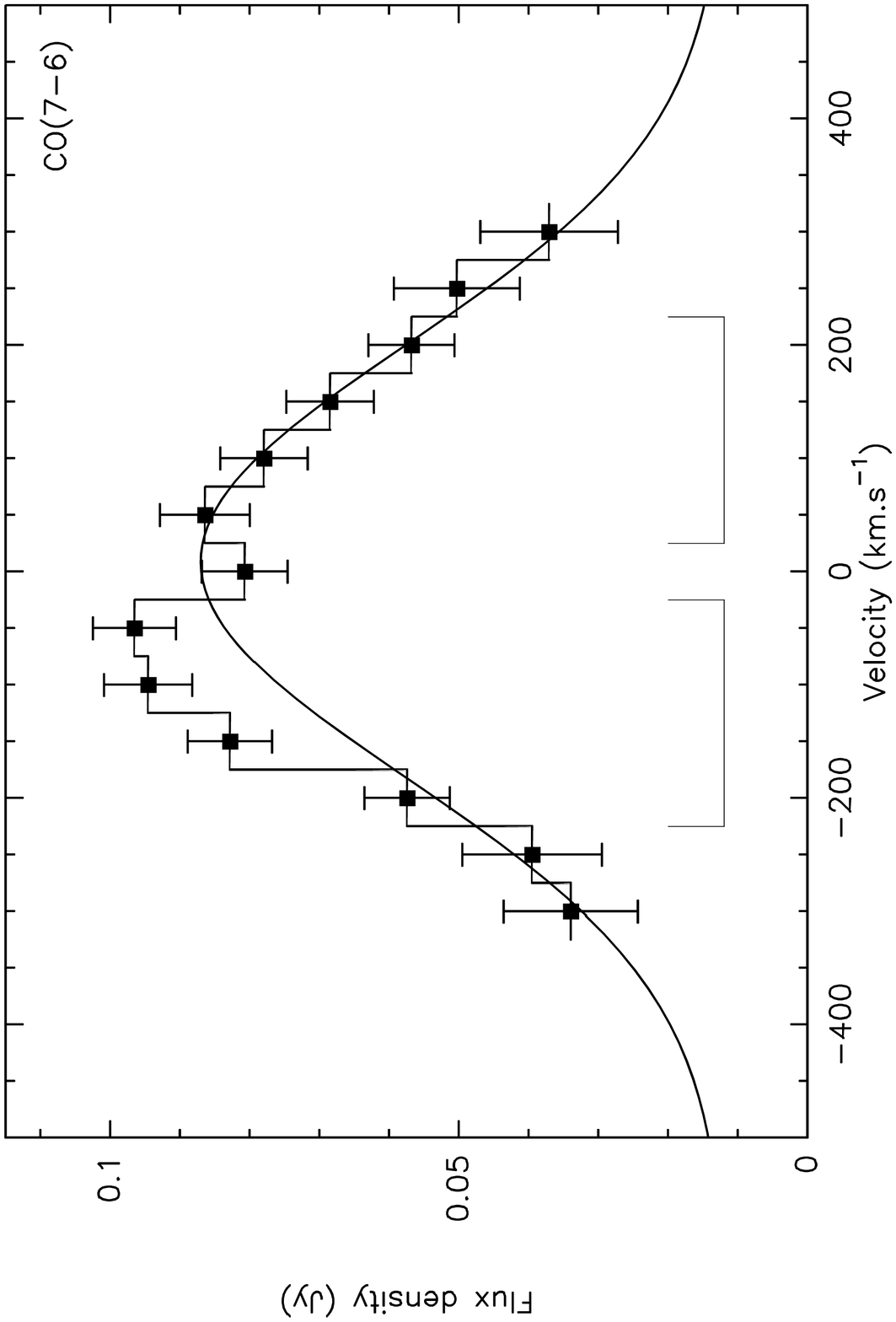},width=8.8cm,angle=270}
\caption{
Spectrum of the CO(7-6) line towards the Cloverleaf. Error bars are
$\pm$ 1$\sigma$. The thin line corresponds to the best gaussian fit to
the wings and the red side with
FWHM$\sim$450 km/s, although it is obvious that the line profile differs from
a smooth gaussian shape and shows an excess of emission on its blue
side. 
The offset of the baseline from zero corresponds to 13 mJy of continuum flux; this is larger than the $5\pm 3$ mJy reported by
Alloin et al (1997), owing the inclusion of shorter baselines in the new dataset and detection of a more extended continuum component.
The central frequency is 226.715 Ghz.
}
\label{fig:lineprofile}
\end{figure}

Our previous CO(7-6) interferometer measurements (Alloin et al 1997)
have been complemented with observations at intermediate baselines,
performed on 1997 April 10 and 21 under excellent conditions.
Calibrations were applied similarly to the previous data set. The
combined data lead to the CLEANed integrated map 
restored with a $0.5''$ circular beam shown in Figure~\ref{fig:iram}a.
In order to search for a velocity gradient, we have first
derived the spatially integrated line profile, following the procedure
already discussed in Alloin et al (1997). The new CO(7-6) line profile,
shown in Figure~\ref{fig:lineprofile}, exhibits a marked asymmetry with a
steep rise on its blue side and a slower decrease on its red side.
Excluding the central velocity channel (so that the split in velocity is
symmetric), we have built the blue
(-225,-25 km/s) and the red (+25,+225 km/s) maps displayed in 
Figures~\ref{fig:iram}c and \ref{fig:iram}d respectively. The difference between
the red-shifted and blue-shifted CLEANed maps (Figure~\ref{fig:iram}b)
establishes definitely the presence of a velocity gradient at the
8$\sigma$ level. The implication of these results on the CO source
will be discussed more thoroughly in Sect. 6.

Measurements of the spot characteristics from the CO(7-6) image have
been performed (spot flux ratios, sizes and orientations) through a
fitting procedure in the visibility domain, as explained in Alloin et al
(1997).
The final parameters are provided in Table~\ref{tab:spotiram} where the 
spot sizes are intrinsic to the image, {\em i.e.} deconvolved by the
interferometer beam.
Although the measurements have not been corrected for seeing effects
(mean seeing estimate of the order of $0.2''$) the spots A, B and C appear
to be definitely elongated (see Figure~\ref{fig:cloverleaf} for spot labels).

\section{HST/WFPC2 observations}

\subsection{The HST/WFPC2 data set}

The HST data of the Cloverleaf used in this analysis have been 
provided by the 
ESO/ST-ECF Science Archive Facility (Garching).
Two data sets obtained with the WFPC2 are presented here. 
The first one was obtained by Turnshek (1994 July 12 [ID:5442] \&
1994 December 23 [ID:5621]) and the second one by Westphal 
(1995 April 27 [ID:5772]). 
Turnshek's observations aimed at discovering the lensing galaxy. They
consist in WFPC-2 images and
FOS spectra of the four quasar images and of the suspected lensing object. 
These observations were not successful
in finding the lensing galaxy, but the FOS spectra of the four quasar spots
have been discussed by Turnshek (1996). 
A recent preprint by Turnshek et al (1997) discussed the PC observations
(ID:5442 and 5621 as well as archival pre-costar observations),
giving astrometry and color variation of the different spots 
(the latter which they explained by dust absorption). They also put some limits
on the detection of the primary lensing galaxy.
Observations of the Cloverleaf by Westphal are part of
a larger programme to study multiple quasars and their environments.
Standard reduction procedures using IRAF/STSDAS packages
have been applied. The absolute photometry was obtained using magnitude
zero-points given in Holtzmann et al. (1995).
Information about the final images in filter 
F336W (central rest wavelength $\sim$ 945 \AA), 
F555W (central rest wavelength $\sim$ 1560 \AA), 
F702W (central rest wavelength $\sim$ 1975 \AA), and 
F814W (central rest wavelength $\sim$ 2290 \AA) are
summarized in Table~\ref{tab:journal}.

\begin{table*}[t]
\begin{tabular}{lcccccc}
\hline
Date & Integration time & Filter & QSO location& Wavelength range & 
contributor & comments \\
YY/MM/DD & seconds & & & (rest frame) \AA & & \\
\hline
94/07/12 & 2400 & F255W  & PC1 & 646--843 & below Lyman limit & No detection \\
94/07/12 & 200 & F702W  & PC1 &1686--2360 & Cont. + CIII] & OK \\
94/07/12 & 2700 & F702W  & PC1 &1686--2360 & Cont. + CIII] & quasar saturated
\\
94/12/23 & 2700 & F336W & PC1 &871--984 & Continuum & PSF not circular \\
94/12/23& 400 & F702W  & PC1 &1686--2360 & Cont. + CIII] & PSF not circular\\
94/12/23 & 900 & F814W & PC1 &1967--2726 & Cont. & PSF not circular\\
95/04/27& 23 & F555W & PC1 &1264--1826 & Cont. + CIV & short exposure\\
95/04/27& 60 & F814W & PC1 &1967--2726 & Cont. & short exposure\\
95/04/27 & 2800 & F814W & WF3 &1967--2726 & Cont. & quasar in WF3 chip\\
\hline
\end{tabular}
\caption{Summary of the HST data sets. Note: the data set of 1994
December shows a weird PSF, though the FGS status was on the FINE
guiding mode: it may be due to some guiding problem resulting from the 
paucity
of guide stars in this area.}
\label{tab:journal}
\end{table*}

\subsection{Properties of the Cloverleaf images}

Two types of information are relevant for our present goal of
modelling the gravitational lens : the relative astrometry (with respect
to spot A, as shown on Figure~\ref{fig:cloverleaf}) as well as the shape of 
the four spots and their intensity ratios.

\begin{figure}
\psfig{figure={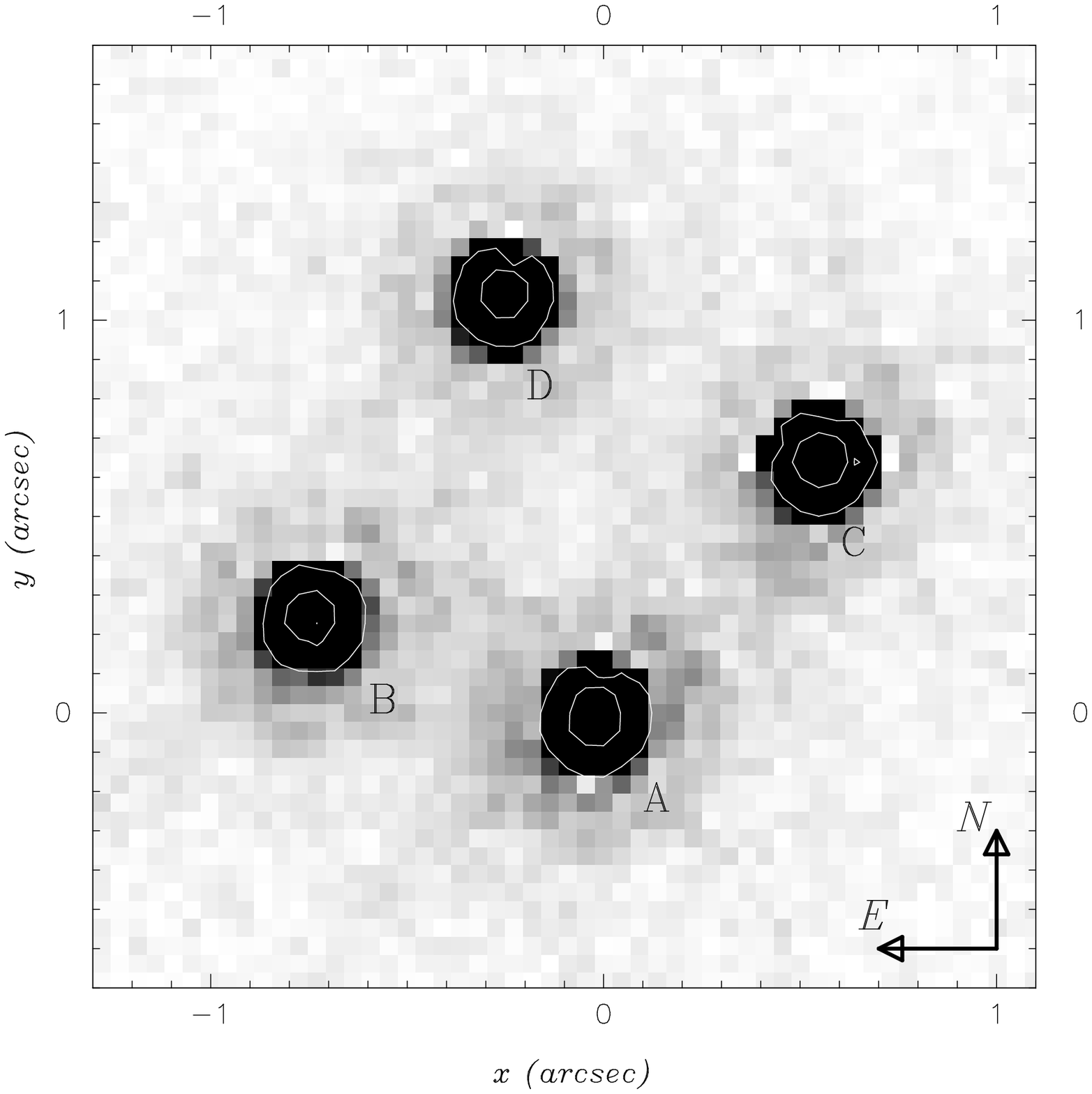},width=8.8cm}
\caption{Image of the Cloverleaf obtained with the HST/WFPC2
through the F702W filter (central rest
wavelength $\sim$ 1975 \AA). The spots shape is perfectly circular.
The spots are labelled A, B, C, D, following Magain et al (1988).
}
\label{fig:cloverleaf}
\end{figure}

Regarding the relative astrometry and spot sizes, the
PC observations in filters F336W, F555W, F702W and F814W
provide similar results displayed in Table~\ref{tab:astrometry}. 
The four spots are stellar-like with a FWHM $\approx$ 1.5 
pixel or $0.068''$.  

\begin{table*}[t]
\begin{tabular}{lcccc}
\hline
Spot & $\alpha$ (J2000) & $\delta$ (J2000) &
$\Delta \alpha$ (arcsec) & $\Delta \delta$  (arcsec)\\
\hline
A & 14 15 46.21 & 11 29 42.75 & 0. & 0.  \\
B & 14 15 46.26 & 11 29 42.92 & +0.75 $\pm$0.01 & +0.17 $\pm$0.01  \\
C & 14 15 45.18 & 11 29 43.48 & -0.49 $\pm$0.01 & +0.73 $\pm$0.01  \\
D & 14 15 46.23 & 11 29 43.79 & +0.36 $\pm$0.01 & +1.04 $\pm$0.01  \\
\hline
\end{tabular}
\caption{Absolute and relative astrometry of the four quasar spots. The
absolute position of the brightest spot A comes from CFHT/FOCAM images
(coordinates are given within a $0.15''$ rms accuracy), while 
the relative positions of B, C and D with respect to A are 
from the WFPC2/PC1 data.
}
\label{tab:astrometry}
\end{table*}

The absolute photometry and the relative intensity ratios (Table~
\ref{tab:photometry}) have been computed with the Sextractor
software (Bertin \& Arnouts 1996) using total and isophotal magnitudes
when images of the quasar were not saturated. No PSF fitting was applied 
as the four different spots are well separated on the PC; furthermore
a PSF fitting  would not have been possible for the December 1994 data
as the PSF was not circular for this observation (see caption of 
Table~\ref{tab:journal}). The important variation of 
the intensity ratios in U compared to V, R and I band
could probably be only  explained by absorption along the line of sight by
intervening galaxies (\HI clouds at redshift $\sim 1.7$ will act as very
efficient absorbers in the U band). This absorption has to be larger
for the spots A \& B which display the biggest relative
change to the C \& D spots. Turnshek et al (1997) proposed that dust
extinction can be the explanation with a preference
for an SMC-like dust extinction at the redshift of the quasar.  
No obvious evidences of microlensing  are detected from the HST data set,
though small variations in the quasar spot intensity ratios are consistent with 
the ESO/NOT monitoring of the Cloverleaf (\O stensen et al 1997).

\begin{table*}[t]
\begin{tabular}{lcccccc}
\hline
& U$_{336W}$ [ratio] & V$_{555W}$ [ratio] & R$_{702W}$ [ratio]
&R$_{702W}$ [ratio]& I$_{814W}$ [ratio]& I$_{814W}$ [ratio]\\
Date & 94/12/23 & 95/04/27 & 94/07/12 & 94/12/23 & 94/12/23 & 95/04/27 \\
\hline
A &   18.95 [1.00]  & 18.02 [1.00]  & 17.71 [1.00]   & 
17.68 [1.00]  & 17.45 [1.00]  & 17.35 [1.00]  \\
B &  19.31 [0.73 $\pm$0.01]  & 18.19 [0.86$\pm$0.01] 
& 17.90 [0.85$\pm$0.01] & 17.82 [0.88$\pm$0.01] 
& 17.58 [0.89$\pm$0.01] & 17.47 [0.89$\pm$0.01] \\
C &  18.93 [1.01 $\pm$0.01]  & 18.25 [0.81$\pm$0.01] 
& 18.02 [0.76$\pm$0.01] & 17.95 [0.77$\pm$0.01] 
& 17.73 [0.76$\pm$0.01] & 17.66 [0.76$\pm$0.01] \\
D &  19.13 [0.85 $\pm$0.01]  & 18.36 [0.75$\pm$0.02] 
& 18.12 [0.69$\pm$0.01] & 17.97 [0.74$\pm$0.01] 
& 17.81 [0.70$\pm$0.01] & 17.73 [0.69$\pm$0.01] \\
\hline
\end{tabular}
\begin{tabular}{lccccc}
& B (K90) & R (Magain et al) & R (K90) & R (Angonin et al) & I (K90) \\
Date& 88/04/27 & 88/03/08 & 88/04/27 & 89/03/07  & 88/04/27\\
\hline
A& [1.00] & [1.00] & [1.00] & [1.00] & [1.00] \\
B& [0.84] & [0.87] & [0.85] &[0.88$\pm$0.02] & [0.90] \\
C& [0.83] & [0.76] & [0.74] &[0.78$\pm$0.02] & [0.76] \\
D& [0.61] & [0.69] & [0.61] &[0.66$\pm$0.02] & [0.59] \\
\hline
\end{tabular}
\caption{Total magnitude and relative intensity ratios 
of the four quasar spots from the WFPC2/PC1 data, compared to previous
observations found in the literature.
}
\label{tab:photometry}
\end{table*}

\subsection{The Cloverleaf environment}

\begin{figure*}[t]
\centerline{
}
\caption{Image of the whole HST field around the Cloverleaf obtained 
with the  WFPC2 camera. The four quasar spots are not resolved in this figure
but are located at the center of the WF3 chip (bottom-right).
The galaxies detected in the field by the SExtractor software
(Bertin \& Arnouts 1996) are overlayed with elliptical contours indicating 
their centroid, orientation and ellipticity. The white
contours are iso-number density of galaxies with $23<I<25$ (ranging from
30 to 80 galaxies/arcmin$^2$). A significant density
enhancement is clearly visible around the Cloverleaf, and is found to be 
almost centered on the 4 quasar spots. This is a good indication that a 
distant cluster of galaxies lies on the line of sight to the 
Cloverleaf: the presence of this cluster certainly  increases the convergence 
of the lensing galaxy.
 }
\label{fig:mosaicI}
\end{figure*}

\begin{figure*}[t]
\centerline{
}
\caption{
F702W mosaic image of the field around the Cloverleaf. The four-spot quasar
is  located at the center of the PC chip. Same convention as in 
Figure~\ref{fig:mosaicI}. The white contours are iso-number densities of
galaxies with $23<R<25$ ranging from 30 to 60 galaxies/arcmin$^2$.
}
\label{fig:mosaicR}
\end{figure*}

Figure~\ref{fig:mosaicI} shows the deep F814W image of the Cloverleaf. 
The four quasar spots lay at the center of the WF3 chip.  
The presence of numerous faint 
objects over a $40''$ region in the environment of the Cloverleaf is striking.  
In order to quantify this effect we have computed the object number 
density, $<$N$>$, and its dispersion, $\sigma_N$, 
in the magnitude range I=23 to 25. We have estimated this density  
in different regions across  the image: in a $40''$ diameter region  
around the Cloverleaf where the
density contrast is clearly visible by eye and in various randomly selected 
areas of similar size.  On the whole frame we find a mean value   
$<$N$>_{whole}=45 \pm 20$ objects/arcmin$^2$ 
(1 $\sigma$). The region around the Cloverleaf has $<$N$>_{Cl}=85 \pm 30 $
objects/arcmin$^2$, with a  peak at 130  objects/arcmin$^2$.   The detection
of a peak is therefore significant at a 4$\sigma$ level. This 
brings us to suspect the presence of a cluster of galaxies
(of unknown redshift), which could contribute significantly to the 
gravitational lens effect. 
We provide in Figure~\ref{fig:mosaicR} the deep F702W image of the
Cloverleaf where the four quasar spots lay at the center of the PC1
chip. Similarly to the F814W image we have plotted the number density
isocontours of faint objects between R=23 to 25. There is no
overdensity larger than 2$\sigma$. Therefore, we conclude that the
4$\sigma$ overdensity around the Cloverleaf is significant.
Fig. \ref{fig:zoom} shows a zoom of this area from the deep F814W image: 
the photometry and
color of the related objects (obtained by combining the deep WF and 
PC observations) are given in Table~\ref{tab:galcolor}. The bulk of them are 
red objects with $R_{702W}-I_{814W}\sim 0.9$ (from 0.7 to 1.2).
While their morphology cannot be obtained from the HST images, their red
color suggest that we might be dealing with
E/S0 galaxies in a high-redshift cluster. 

\begin{figure}[t]
\centerline{
}
\caption{Zoom of the deep F814W image of the Cloverleaf
(rotated to be in RA-DEC coordinate system: x-axis is -$\alpha$,
y-axis is +$\delta$).
The galaxies with $22.5<I<25$ detected in the field by the 
SExtractor software (Bertin \& Arnouts 1996) are overlayed with 
elliptical contours indicating 
their centroid, orientation and ellipticity. Photometry and color of
these galaxies are provided in Table~\ref{tab:galcolor}.
The dashed square overlaid corresponds to the location of the 
PC R$_{F702W}$ image. The cross correspond to the center chosen for the
cluster mass component (model 2).
}
\label{fig:zoom}
\end{figure}

\begin{table}[t]
\begin{tabular}{lcccc}
\hline
ID Number & $\Delta\alpha$ & $\Delta\delta$ & I$_{814W}$ & R$_{702W}$ - I$_{814W}$  \\
& (arcsec) & (arcsec) &  & \\
\hline
1 & -4.7 & -6.2 & 22.52 & 0.72\\
2 & -5.6 & -3.9 & 23.11 & 0.86\\
3 & 7.3 & -3.7 & 23.16 & 0.88\\
4 & 13.2 & 8.7 & 23.57 & 0.59\\
5 & -13.7 & 16.2 & 23.61 & 1.01\\
6 & 7.2 & 13.8 & 23.73 & 0.94\\
7 & -3.9 & 5.4 & 23.75 & 0.74\\
8 & -4.3 & 0.6 & 23.78 & 1.16\\
9 & 0.1 & -5.8 & 23.94 & 0.12\\
10 & -14.7 & 9.1 & 23.95 & 1.09\\
11 & -5.4 & 5.3 & 24.01 & 1.09\\
12 & -14.2 & -2.4 & 24.20 & 0.43\\
13 & 8.0 & -11.6 & 24.20 & 0.92\\
14 & 1.7 & 4.2 & 24.37 & $>$ 1.1\\
15 & -8.2 & 0.6 & 24.41 & 1.05\\
16 & 11.4 & -6.7 & 24.55 & $>$ 0.9\\
17 & 3.4 & 0.1 & 24.62 & $>$ 0.9\\
18 & -7.2 & 14.4 & 24.66 & 0.91\\
19 & 4.5 & -2.9 & 24.74 & 0.88\\
20 & -8.1 & -3.6 & 24.78 & $>$ 0.7\\
21 & -2.7 & -14.4 & 24.80 & $>$ 0.7\\
22 & -9.7 & 12.9 & 24.84 & $>$ 0.7\\
23 & 2.2 & -9.1 & 24.89 & $>$ 0.6\\
24 & 12.0 & 0.3 & 24.93 & 0.13\\
\hline
\end{tabular}
\caption{Relative position from the quasar spot A, 
total magnitude and color of the galaxies detected in I$_{F814W}$
(numbered as in Figure~\ref{fig:zoom}).
Typical error for the faintest object is 0.2 mag.
}
\label{tab:galcolor}
\end{table}

The small size (of the order of $0.3''$ which hampers  any
strong morphological classification), faint magnitude ($23<I$) and
red color $R_{702W}-I_{814W}\sim 0.9$ of these galaxies 
are suggestive  of  {\sl high-redshift }  E/S0 galaxies
(as can be found for example in the CFRS  
sub-sample of galaxies with 1$<$z$<$2
(Lilly et al 1995)). Moreover, the significant {\em concentration}
of these galaxies
near the particular line of sight to the multiple quasar
is a  convincing indicator of the presence of a
{\sl high-redshift cluster}.
This cluster could either be linked physically to the quasar 
at redshift z$\sim$2.56 or
along the line of sight.
Hereafter, we {\em assume} this cluster to be at a redshift close 
to that of the  
narrow absorption systems observed by Turnshek et al (1988),
and  Magain et al (1988): either 1.438, 1.661, 1.87 or 2.07. 
For the sake of simplicity, we shall adopt hereafter a value of 1.7.
Since no observations are publicly available yet in the near infrared, 
it is practically impossible to obtain secure photometric redshifts in this
redshift range ({\it e.g.} the 4000 \AA \ break would fall at 1$\mu$m at z=1.7).
Dedicated high-resolution IR imaging and/or optical/IR 
spectroscopy should be 
carried out on the objects forming this overdensity in order to 
determine their distance and nature.\\

We have examined the alternative possibility that this ensemble of faint sources
corresponds to a population of \HII regions in a nearby galaxy with low 
surface brightness, extending over $40''$ in diameter. 
The intrinsic size of such a dwarf galactic system, around 10 kpc,
requires that its distance should be  less than 50 Mpc
(or its redshift less than 0.01). However, the observed R-I$\sim$ 0.9 color 
of the objects forming the overdensity, is far too red to fit
the R-I color of giant 
extragalactic \HII regions observed in a sample of nearby \HII galaxies,
R-I$\sim$~-0.17 (Telles et al 1997) or as modelled, R-I$\sim$~-0.2
({\it e.g.} Bica et al 1990).
Furthermore, the observed I magnitudes are too faint by at least 7 magnitudes
with respect to expected values of such an \HII population located 
in the local Universe. Therefore, this interpretation should be discarded.

\subsection{The lensing object}

Although the cluster candidate mentioned earlier is acting 
as an additional lensing agent, 
the small angular separation between the four spots implies that most of the 
lensing effect is likely produced by a galaxy located amid the four
spots of the quasar.
This lensing galaxy has not been detected so far on either
``deep" visible and near infrared ground based observations (Angonin et al
1990; Lawrence 1996) nor on the PC observations by Turnshek et al (97). 

Yet, the detection of the lensing galaxy would be of great importance to 
model
accurately the mass distribution of the lens as we could then
assume that the mass follows the geometry of the light, as in other 
successful lensing model calculations and as expected in the inner regions
of galaxies. 

We tentatively used the deepest 
WFPC2/F814W images in order to detect the central lensing galaxy. The four 
spots of
the quasar have been subtracted using the PSF model provided by the
nearby star. Due to the significant light residual 
related to the quasar photon noise, the telescope diffusion and the 
first diffraction rings, 
the limiting surface brightness in the center of the Cloverleaf is at least
1 mag lower than across the whole image. This reduces the limiting
magnitude to a  value of I$\sim$24, though the faintest objects
detected in the field have I$\sim$25.  Even at such a faint magnitude, there is
no evidence of a signal spread over few pixels.  We conclude that the
lensing galaxy must have $I>24$ . We can set an upper
limit on its velocity dispersion, assuming the extreme case where the
lens is an elliptical galaxy fainter than I=24 with a redshift 
close to that of the narrow absorption lines seen on the quasar spectrum at    
 $z\sim 1.7$.   In order to
infer the absolute magnitude of the corresponding lens, we have run  
 a  galaxy evolution model for an elliptical galaxy with 
standard parameters
 (power law Initial Mass Function with one segment
ranging from 0.1 to 125 M$_{\odot}$, single burst of star
formation, $\Omega$=1., $H_0$=50 km/sec/Mpc)
from Bruzual \& Charlot (1993). We find that in the most favorable case,
the galaxy should have an absolute magnitude larger than $M=-21.3$ in 
order not to be detected.  Using the
Faber-Jackson relation (Fall 1981), we derive an upper limit of 
its velocity
dispersion  $\sigma=350$ km/sec.  A similar study can be performed 
using the 2000 sec. K-band exposure obtained at Keck (Lawrence 1996).
Given the expected throughput of the Keck camera, we estimate the central 
lensing galaxy to have $K>22$, which is a constraint not as stringent 
as that derived from the HST images. Hence, the visible and near
infrared data are not sufficient to constrain the mass of the lensing galaxy.
But in fact, if we take into account the likely presence of the distant 
cluster, it is
no longer  necessary  to invoke a large mass (i.e. high brightness) for the  
lensing galaxy since a fraction of the convergence and the
deflection angle would be contributed by the cluster.

\section{Additional constraints from the absolute registration of images
in various wavebands}

In order to derive precisely the shear induced by the gravitational lens
on an extended source in the quasar, it is imperative to register with
a high accuracy the Cloverleaf image in a waveband corresponding to 
  point-like images of the quasar (i.e. R-band corresponding to rest
wavelength 1967 \AA) and in a waveband corresponding to an extended
source in the quasar (i.e. the CO(7-6) molecular emission).  The high
precision required, better than $0.2''$, cannot be achieved from the HST data
because of their limited field of view. Hence, we have searched for 
wide-field 
exposures of the Cloverleaf field obtained under good seeing
conditions at CFHT.

\subsection{The FOCAM CFHT data set}

We have retrieved from the CFHT data archive (CADC database in Victoria) 
three
images obtained on 1992 February 27 by Angonin, Vanderriest and
Chatzichristou, with FOCAM.  The CCD in use was Lick II (1500 pixels of
$0.2''$), covering a $5'$ field. Two images were obtained through the R band
and one through the I band, under a $0.6''$ seeing condition. 
For every image, the integration time was
310 seconds. To derive the astrometry, we used the two R band direct
images, unprocessed since we are only interested in the astrometric
information.

\subsection{Absolute astrometry}

The absolute astrometry was performed using 6 reference stars across the
field, from the Cambridge APM database (much deeper than the
Guide Star catalogue). This 
allows  absolute positioning of the brightest spot A within an
accuracy of $0.15''$ (rms). Then, we have used the relative astrometry of
spots B, C and D, with respect to A, from the HST image  (relative
accuracy $\pm 0.01''$).  The absolute astrometry of the Cloverleaf 
is provided in Table~\ref{tab:astrometry} (J2000.0).

\subsection{Registration of the optical and millimeter images}

The IRAM CO(7-6) image is obtained within an absolute
astrometric accuracy of $0.1''$ (rms), quite comparable to that achieved from 
the combined CFHT/FOCAM and HST/WFPC2 results.  
Therefore, it is possible to register both the optical and the
millimeter data in an absolute manner (Figure~\ref{fig:iram}).
This is the starting point  of our new modelling of the Cloverleaf.

In summary, we
consider for the modelling  the positions as well as R and I intensity
ratios (the least affected by dust absorption) 
of the four spots given in  Table~\ref{tab:photometry}, 
since we know that they are not affected by any spatial resolution 
effect. Conversely, the relative positions of the CO spots with respect to the
visible spots are subject to some spatial resolution effect because 
the interferometer natural beam size is of the order of the spot size. 
Therefore, we shall not use the CO spot positions to build the lens
model, but rather check whether the CO spots predicted by the lens model 
and convolved by the interferometer beam reproduce the observed 
offsets between the CO and the visible spots (see Sect. 5 and 6)

\section{New model of the Cloverleaf gravitational lens}

\subsection{The procedure}
The modelling of the Cloverleaf is based on the minimization algorithm 
described previously in several papers (Kneib et al 1994, 1996) and
used to model giant arcs in cluster-lenses. This algorithm adjusts 
the parameters of the model through a minimization of  the differences 
in the position and the geometry of the  lensed
spots once they are sent back to the source plane. The fitting uses
observational constraints like the position, intensity and shape of the lensed
spots and eventually the light distribution associated with the
deflector as additional information on the mass distribution. The model
incorporates parameters of the lensing potential, through a simple 
analytical representation of the mass distribution. In the present case, we 
use a truncated elliptical mass distribution (EMD) defined as the
difference of two pseudo-isothermal elliptical models (PIEMD)
(Kassiola \& Kovner 1993, Hjorth \& Kneib 1997):
\begin{eqnarray}
\Sigma(R) & = & \Sigma_0 {as \over s-a}
          \left ( {1 \over \sqrt{a^2+R^2}}- {1 \over \sqrt{s^2+R^2}}
\right )
\end{eqnarray}
where $a$ is the core radius and $s$ is the truncature radius.
Moreover we have $R^2= (x-x_0)^2/(1-\varepsilon) +
(y-y_0)^2/(1+\varepsilon)$, where ($x_0,y_0$) are the centre position,
($x,y$) the current position in the principal axis of the
lens, and $\varepsilon=(a^2-b^2)/(a^2+b^2)$ is the ellipticity of the mass
distribution. The important characteristic of such a model is that the
mass
distribution has an elliptical symmetry whatever $\varepsilon$. Its
$1/R^3$ dependency at large radii imposes a finite mass to the model,
and is compatible with the theoretical prescription 
of violent relaxation models (Hjorth \& Madsen 1991).
Furthermore, the model treatment is fully analytical.

The constraints used for the gravitational lens modelling are as
follows:
\begin{enumerate}
\item the {\sl relative positions} of the quasar spots from the HST and the 
{\sl intensity ratio} taken in the R and I band.
These constraints will primarily
enable us to determine the mass model.

\item the {\sl non-detection} of a 5th spot, which puts a limit on the
size of the lens core.

\item the {\sl position of the cluster center}  as measured from the
overdensity  of galaxies near the Cloverleaf. 

\item both the cluster and the lensing galaxy are assumed (for convenience) 
to be at z=1.7.

\end{enumerate} 

The {\sl relative intensity ratios and the measured shapes of the CO 
spots} are used as a test of the model, and provide as well information
on the size and geometry of the CO source.

\subsection{Results}

We have computed two types of model: model 1 which includes an individual
galaxy with a dark halo at z=1.7 and model 2 which considers an
individual galaxy and a cluster both at z=1.7. These models
are not unique but give similar qualitative results.  
The parameters for the lensing galaxy and the cluster component 
are shown in Table~\ref{tab:model}. 
Model 2 is dominated by the shear of the
cluster (as the center of the cluster is close to the Cloverleaf --
PA position of the cluster center is $\sim$35 deg),
which explains why the PA of the lensing galaxy is negative and
different than in the model 1.
The cluster component is poorly
known since we were not able to derive significant constraints on the 
lensing galaxy from the photometry (see Sect. 3.3). Yet, once an upper limit
has been found for the velocity dispersion of the lensing galaxy, any model 
including an additional lens-plane (such as that of the galaxy cluster)
and implying a mass, hence a velocity dispersion for the individual 
lensing galaxy 
lower than this limit, is formally acceptable, provided the gravitational 
shear observed on the CO map is modelled equally well.

\begin{table}[t]
\begin{tabular}{lcc}
\multicolumn{3}{c}{Model 1: $\chi^2$= 1.9}\\
\hline
Positions & Galaxy ($z=1.7$)& Dark Halo ($z=1.7$)  \\
\hline
$\Delta\alpha$ & 0.187$\pm$0.01 & 0.187  \\
$\Delta\delta$ & 0.563$\pm$0.01 & 0.563 \\
$a$ ($h_{50}^{-1}$ kpc) & 0.05 (fixed)& 4.4$\pm$1  \\
$s$ ($h_{50}^{-1}$ kpc) & 20.0 $\pm$5& 250$\pm$20 \\
PA & 21.5 $^o$$\pm$3& 21.5$^o$  \\
$\varepsilon$ & 0.1$\pm$0.05 & 0.2$\pm$0.1 \\
$\sigma$ (km/sec.) & 127$\pm$10  & 395$\pm$20  \\
\hline
\noalign{\medskip}
\multicolumn{3}{c}{Model 2: $\chi^2$= 2.4}\\
\hline
Positions & Galaxy ($z=1.7$)& Cluster ($z=1.7$)  \\
\hline
$\Delta\alpha$ & 0.174$\pm$0.01 & -4.7 (fixed)  \\
$\Delta\delta$ & 0.548$\pm$0.01  & -6.2  (fixed) \\
$a$ ($h_{50}^{-1}$ kpc) & 0.05 (fixed) & 50$\pm$10  \\
$s$ ($h_{50}^{-1}$ kpc) & 30. $\pm$5 & 700 (fixed)\\
PA & -32$^o$ $\pm$2& -30 $\pm$5 \\
$\varepsilon$ & 0.33$\pm$0.05 & 0.2$\pm$0.1 \\
$\sigma$ (km/sec.) & 230 $\pm$10 & 1100 $\pm$100  \\
\hline
\end{tabular}
\caption{Results of the lens modelling of the Cloverleaf.
The first model does not include a cluster component, while the second 
one does. Note that the cluster model can be even more complex
and it should be considered only as a one particular solution.
}
\label{tab:model}
\end{table}

\begin{figure*}

\begin{minipage}{5.5cm}
\psfig{figure={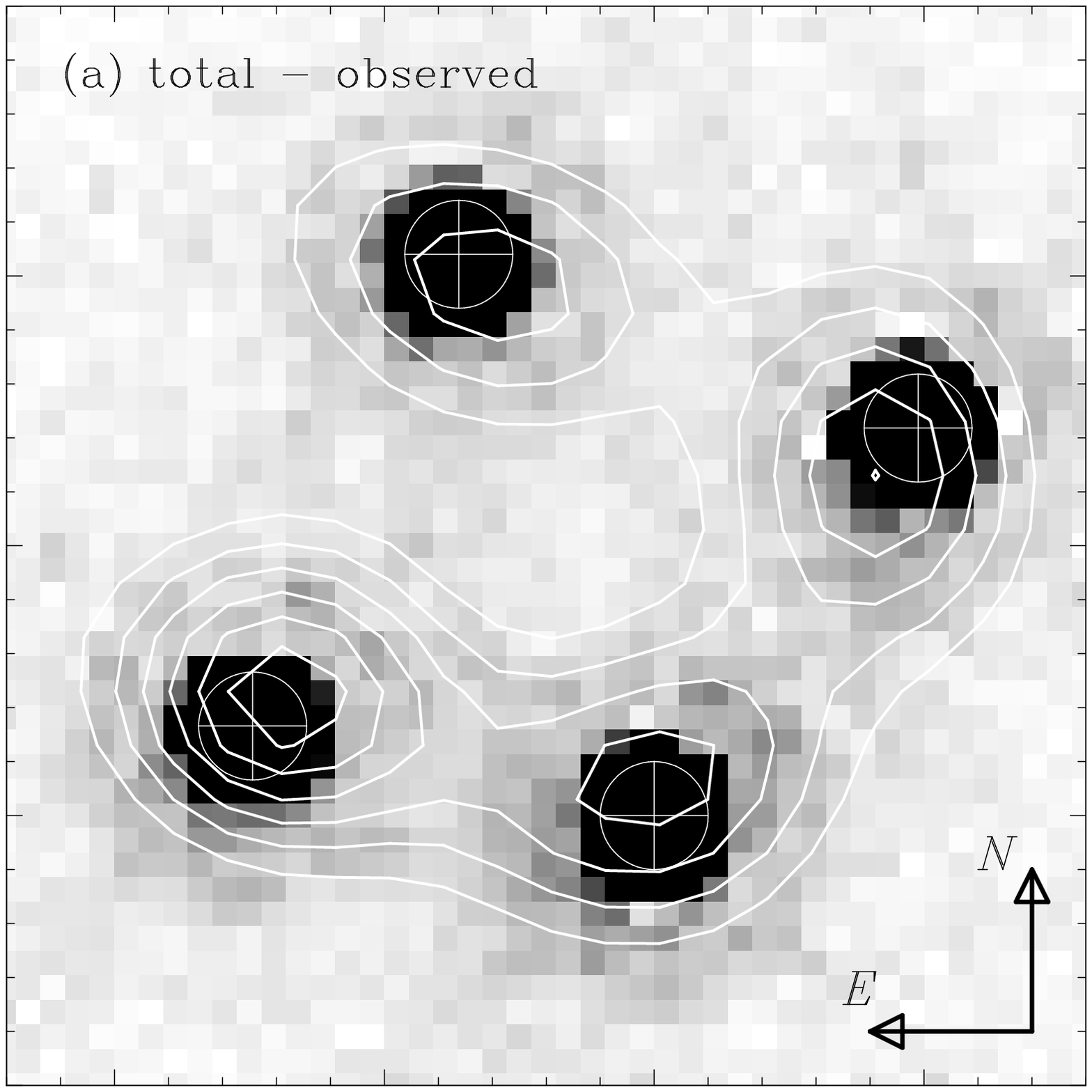},width=5.3cm}
\end{minipage}
\begin{minipage}{5.5cm}
\psfig{figure={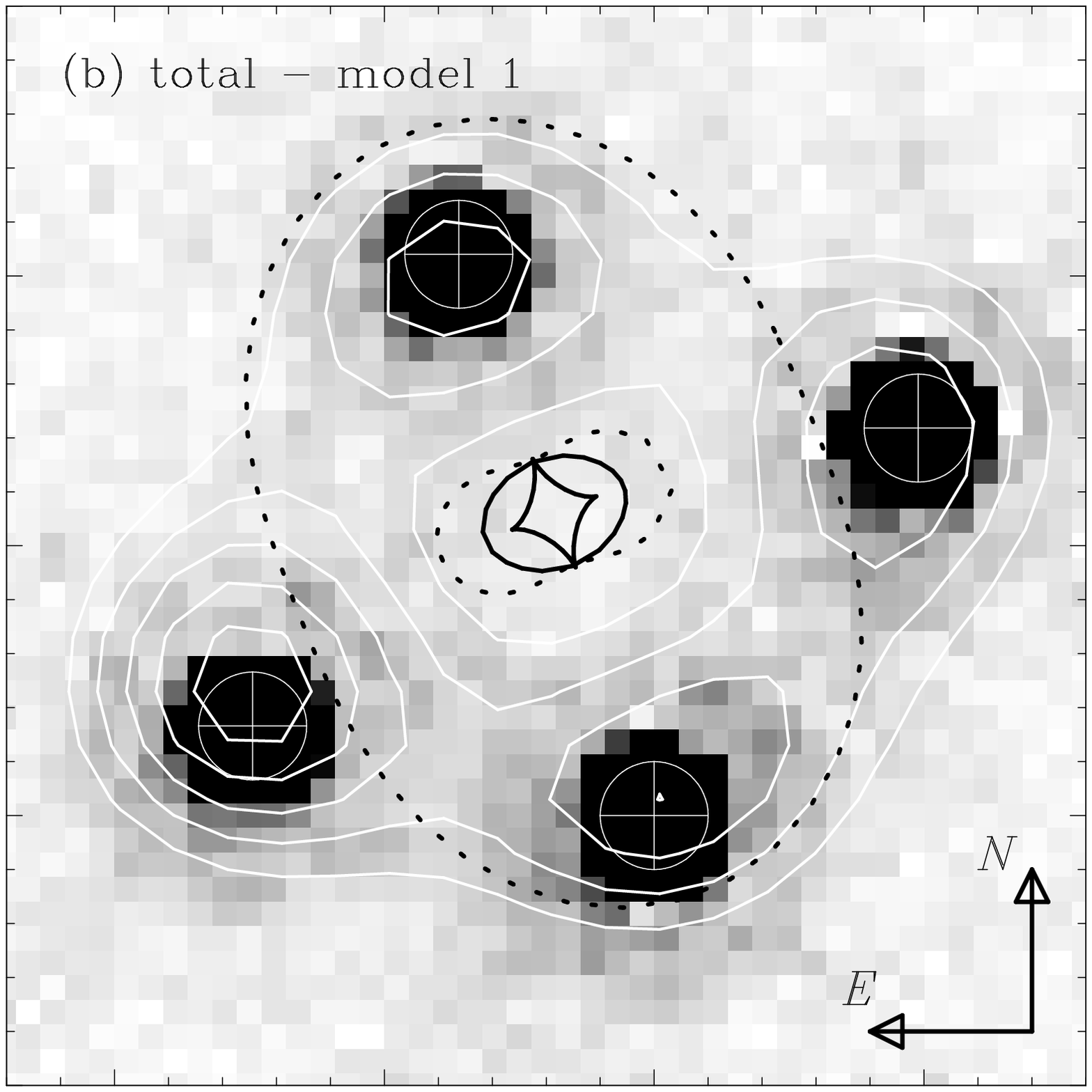},width=5.3cm}
\end{minipage}
\begin{minipage}{5.5cm}
\psfig{figure={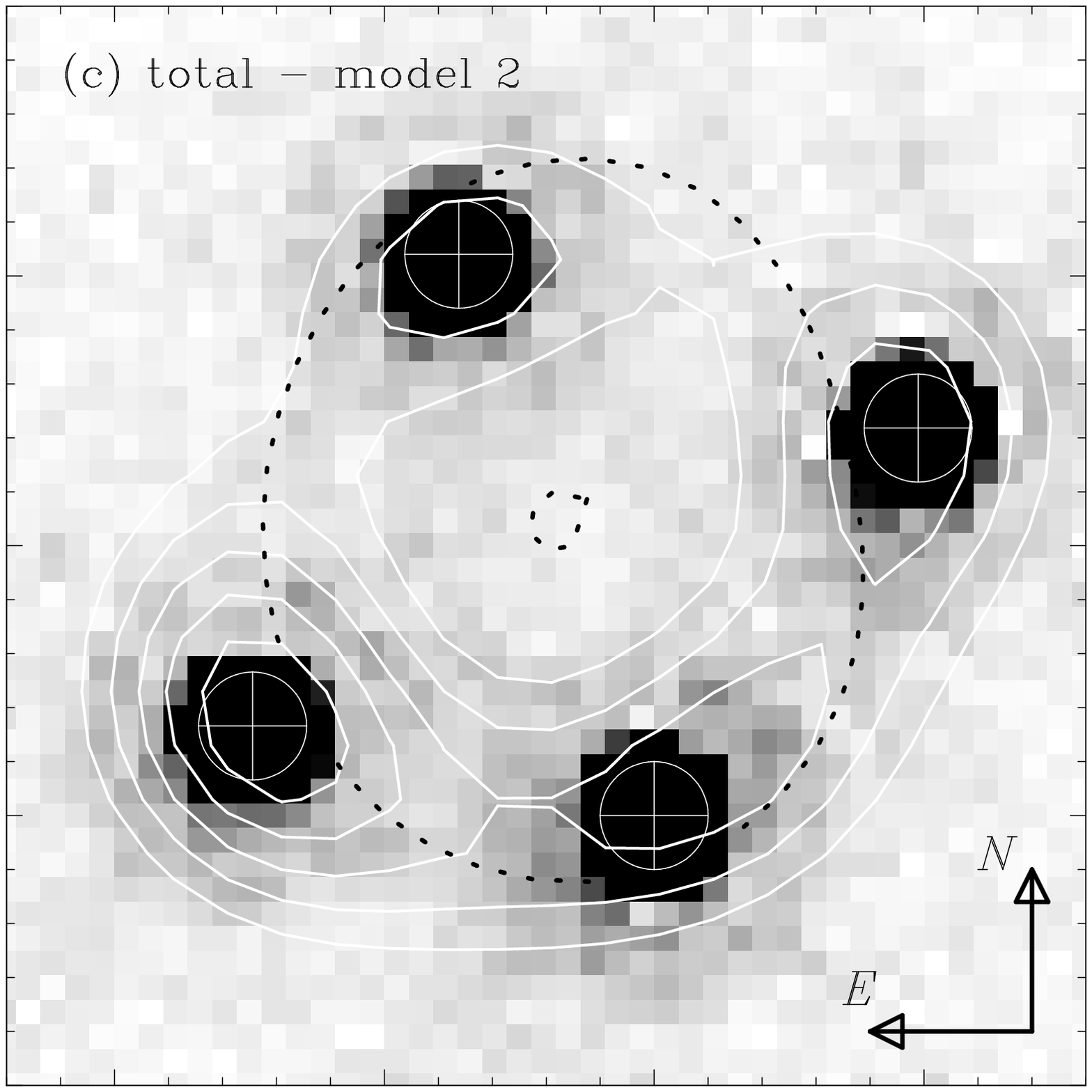},width=5.3cm}
\end{minipage}

\begin{minipage}{5.5cm}
\psfig{figure={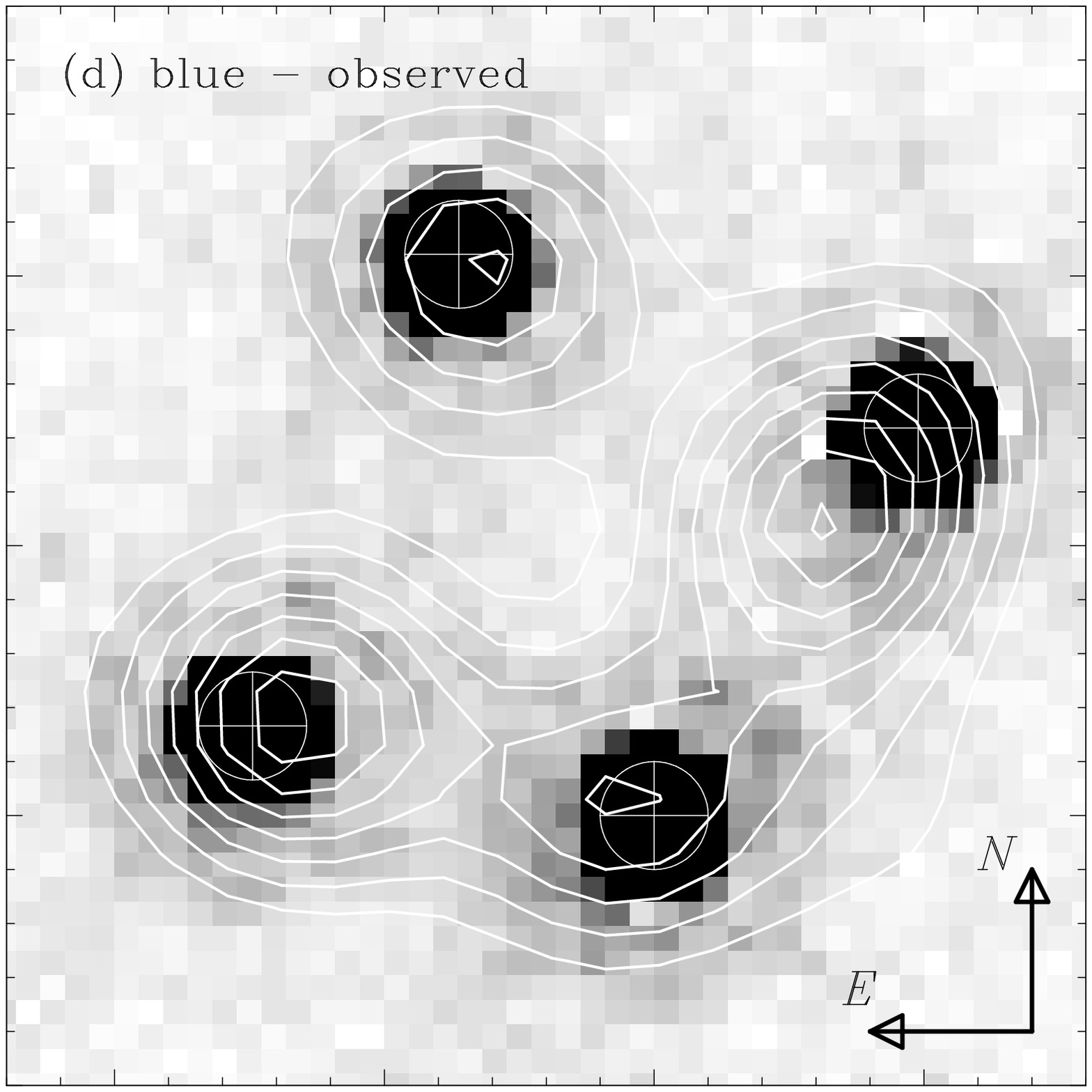},width=5.3cm}
\end{minipage}
\begin{minipage}{5.5cm}
\psfig{figure={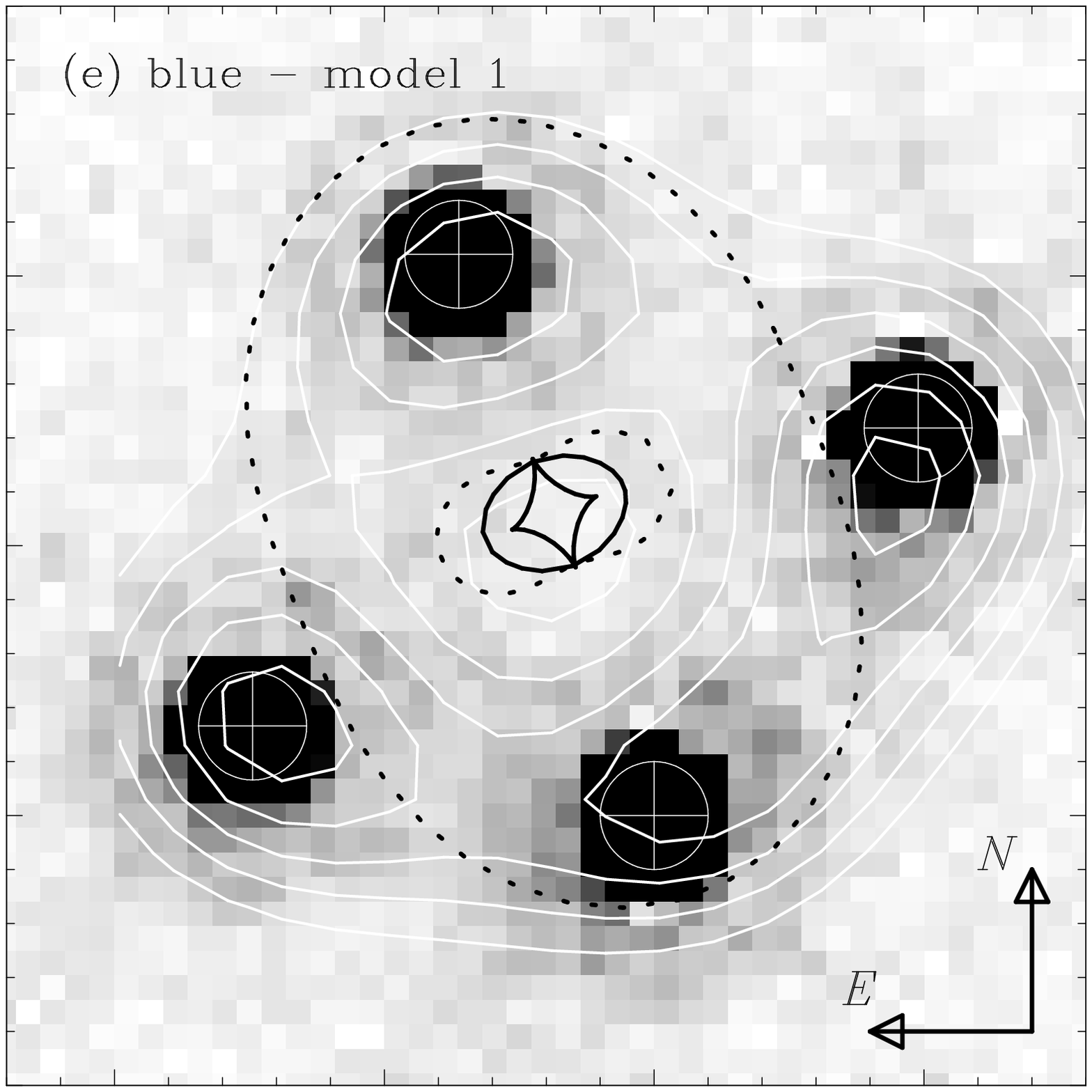},width=5.3cm}
\end{minipage}
\begin{minipage}{5.5cm}
\psfig{figure={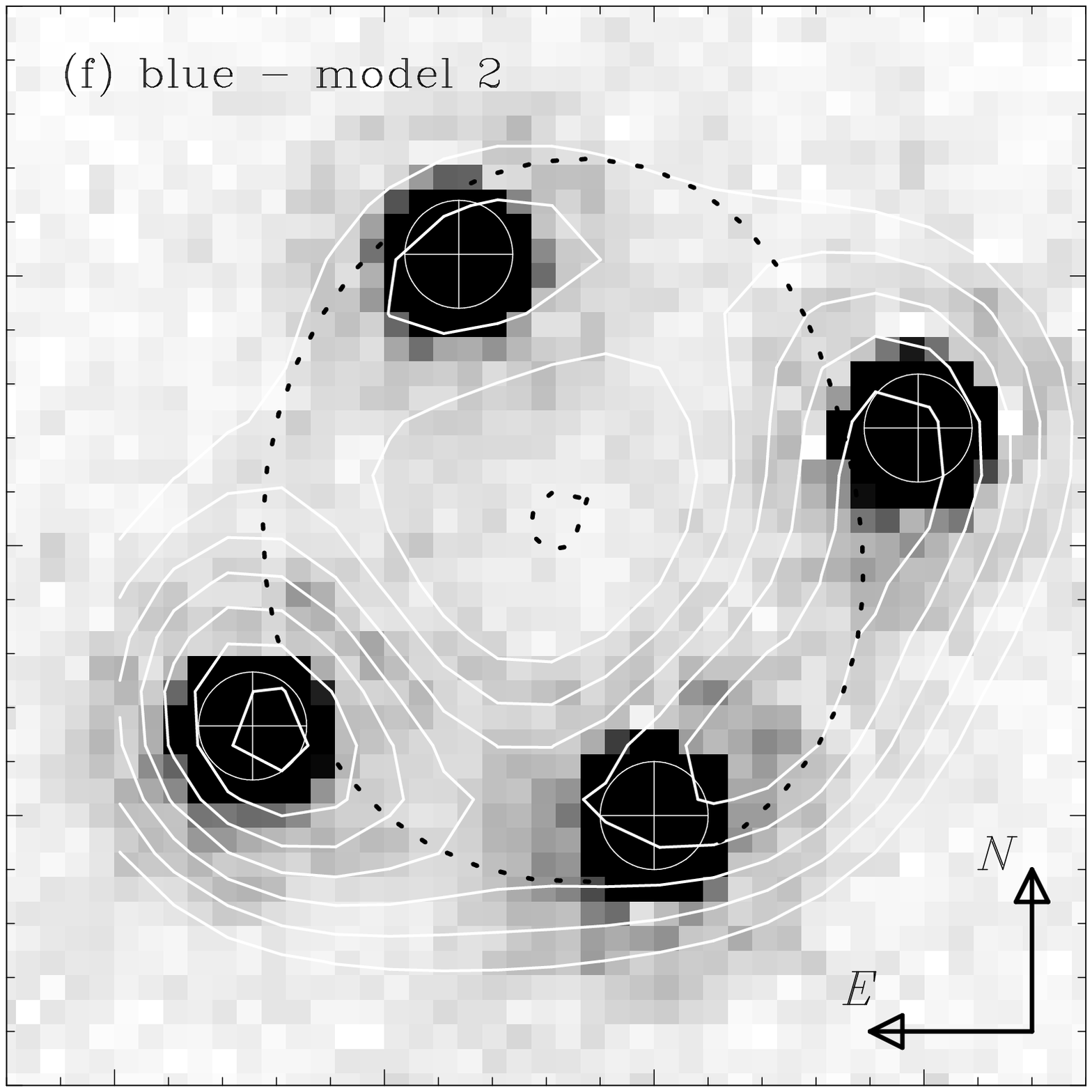},width=5.3cm}
\end{minipage}

\begin{minipage}{5.5cm}
\psfig{figure={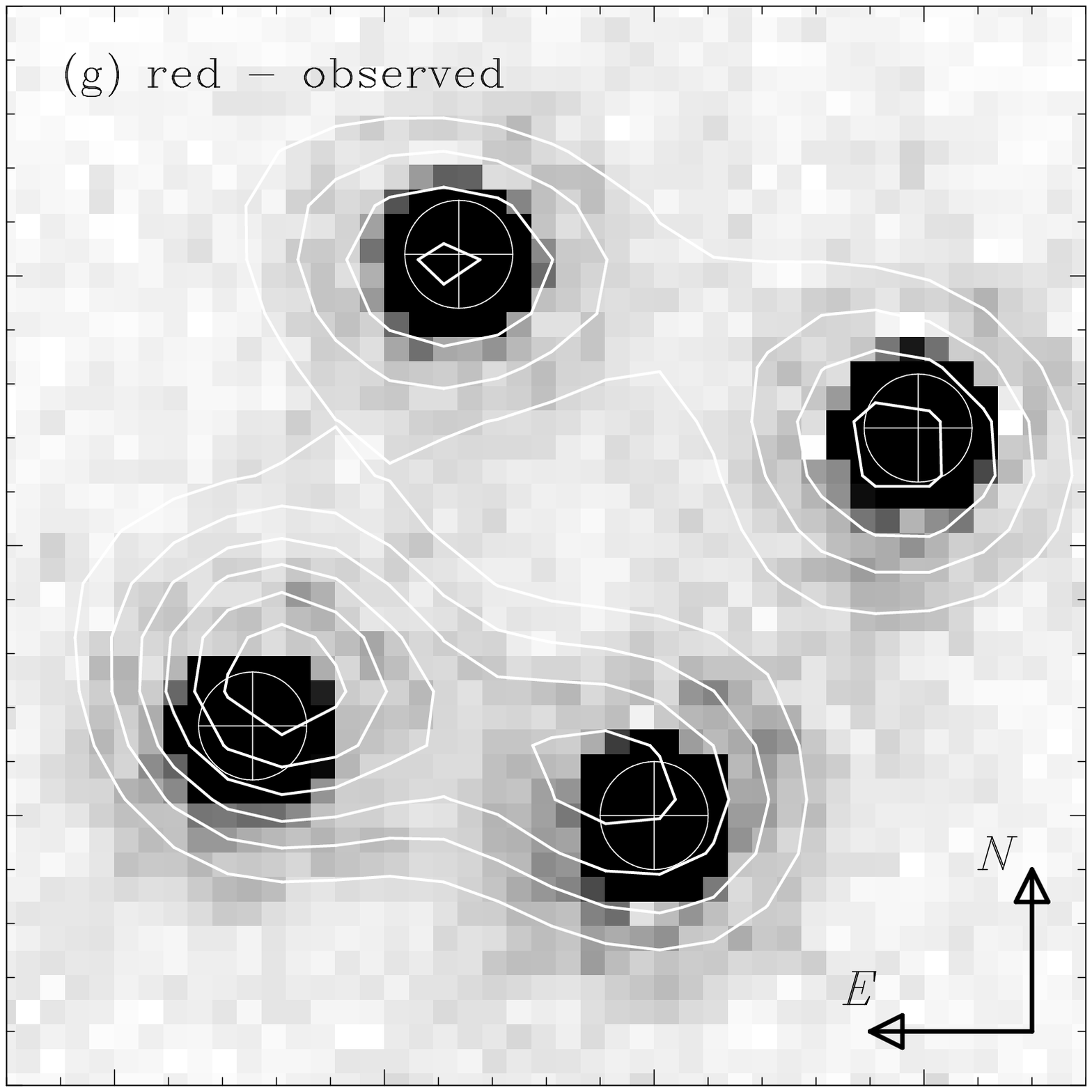},width=5.3cm}
\end{minipage}
\begin{minipage}{5.5cm}
\psfig{figure={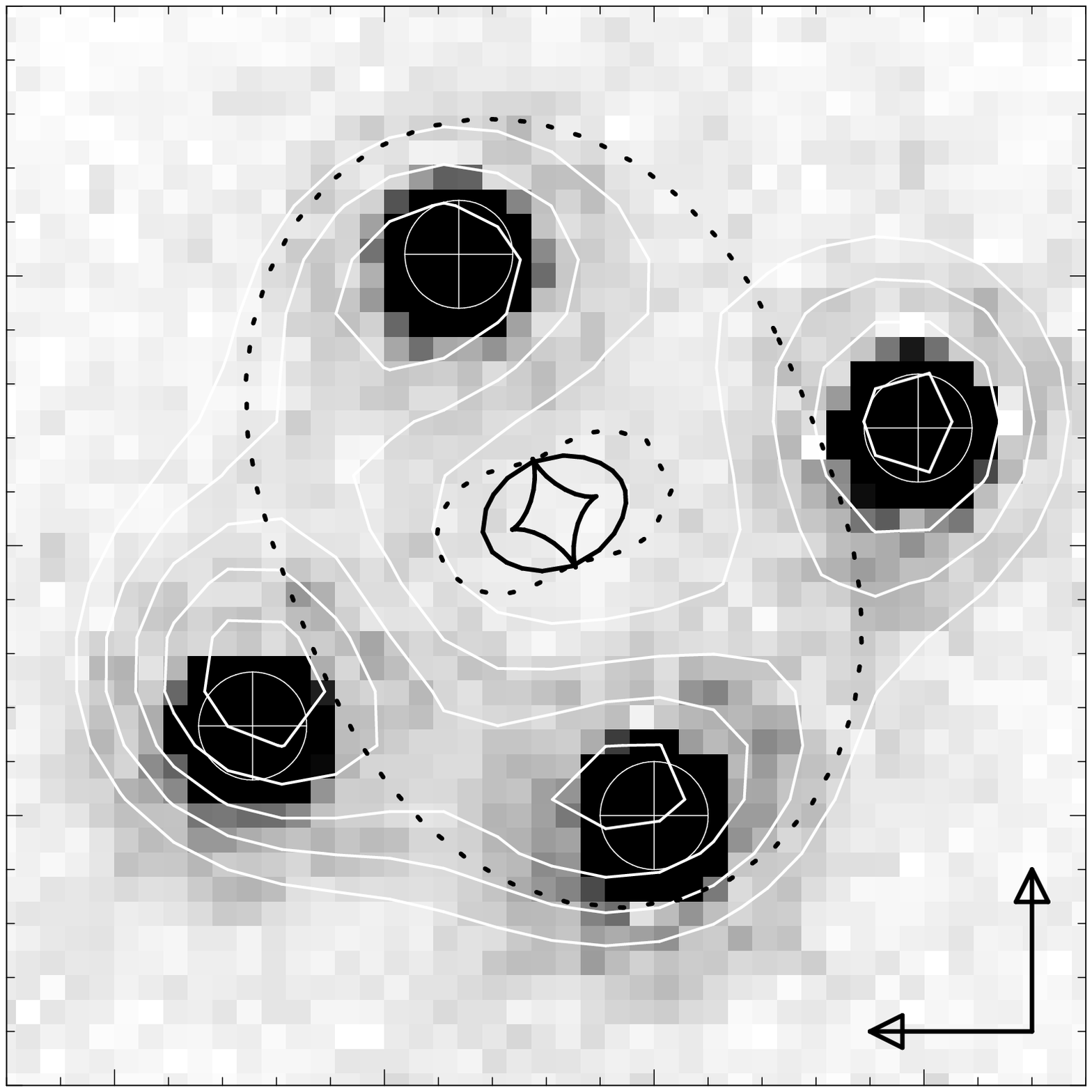},width=5.3cm}
\end{minipage}
\begin{minipage}{5.5cm}
\psfig{figure={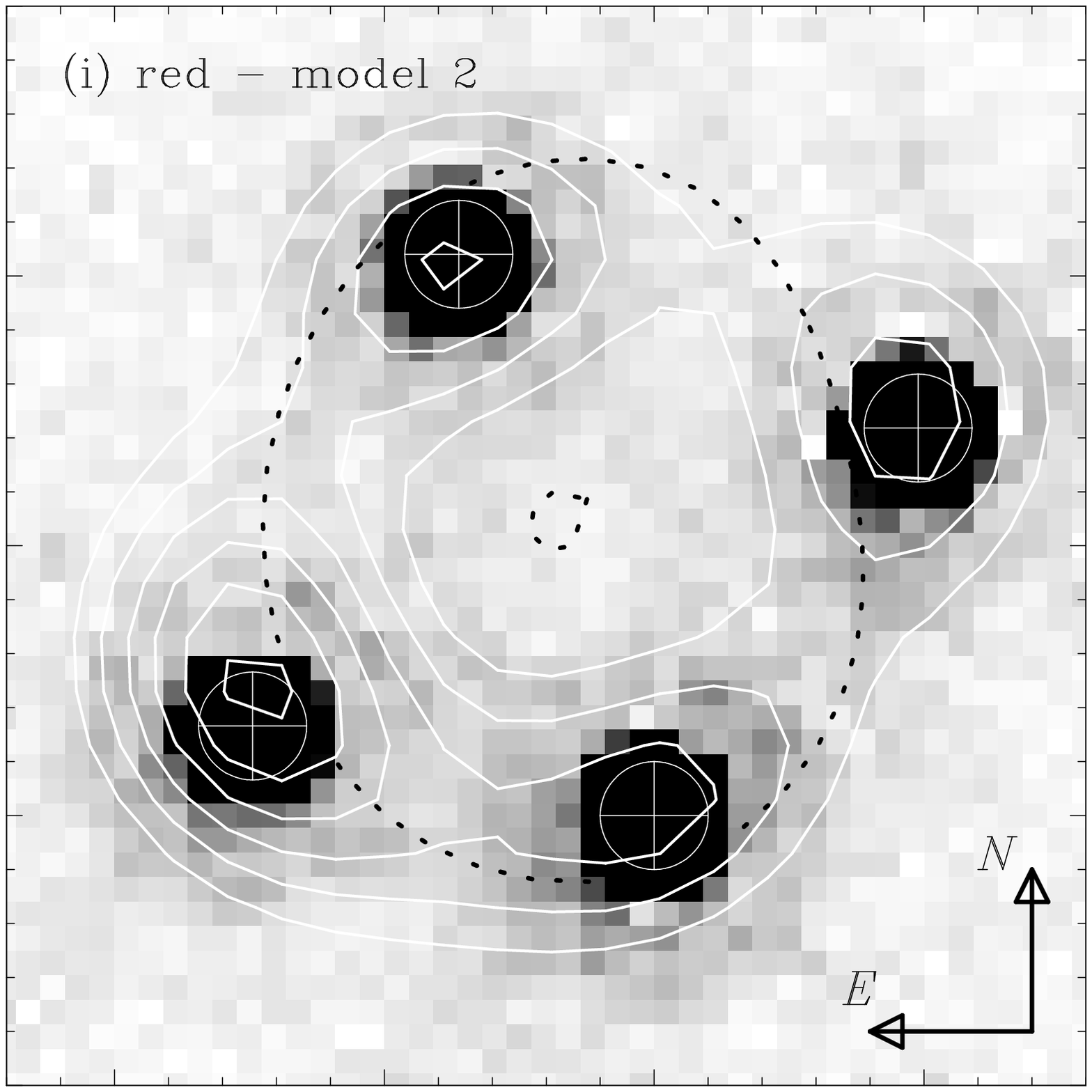},width=5.3cm}
\end{minipage}

\begin{minipage}{5.5cm}
\psfig{figure={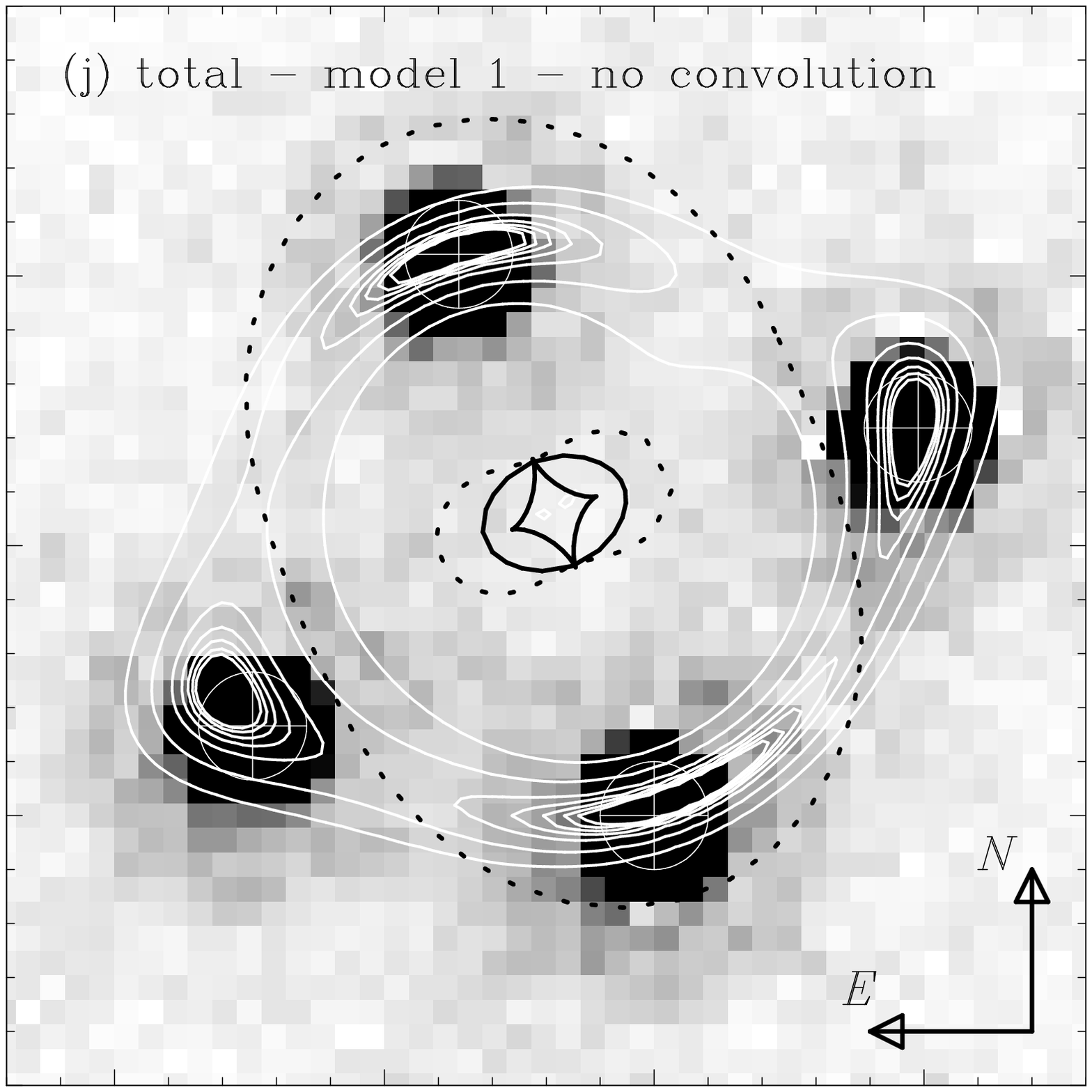},width=5.3cm}
\end{minipage}
\begin{minipage}{5.5cm}
\centerline{\psfig{figure={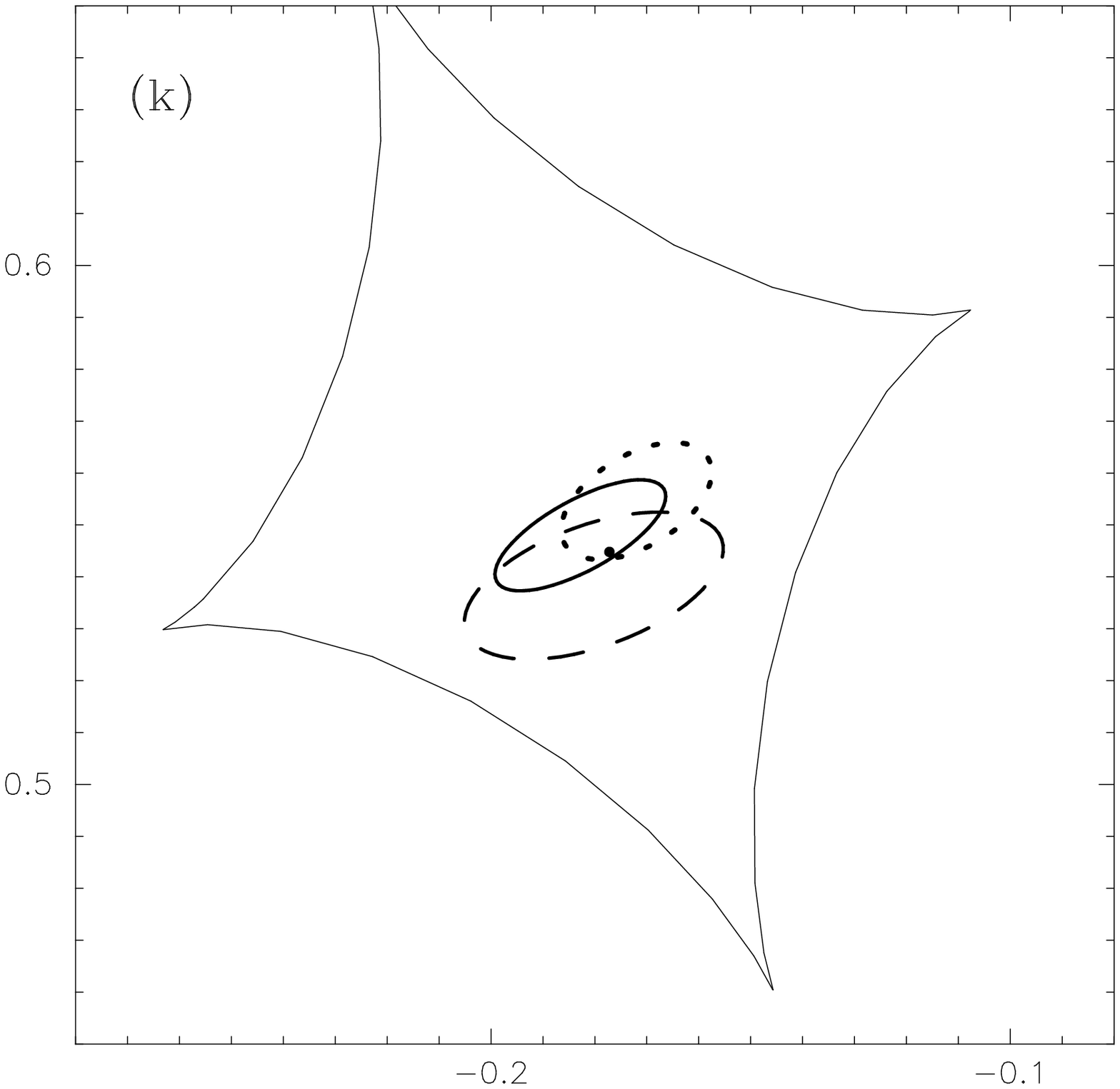},width=5.0cm}}
\end{minipage}
\begin{minipage}{5.5cm}
\centerline{\psfig{figure={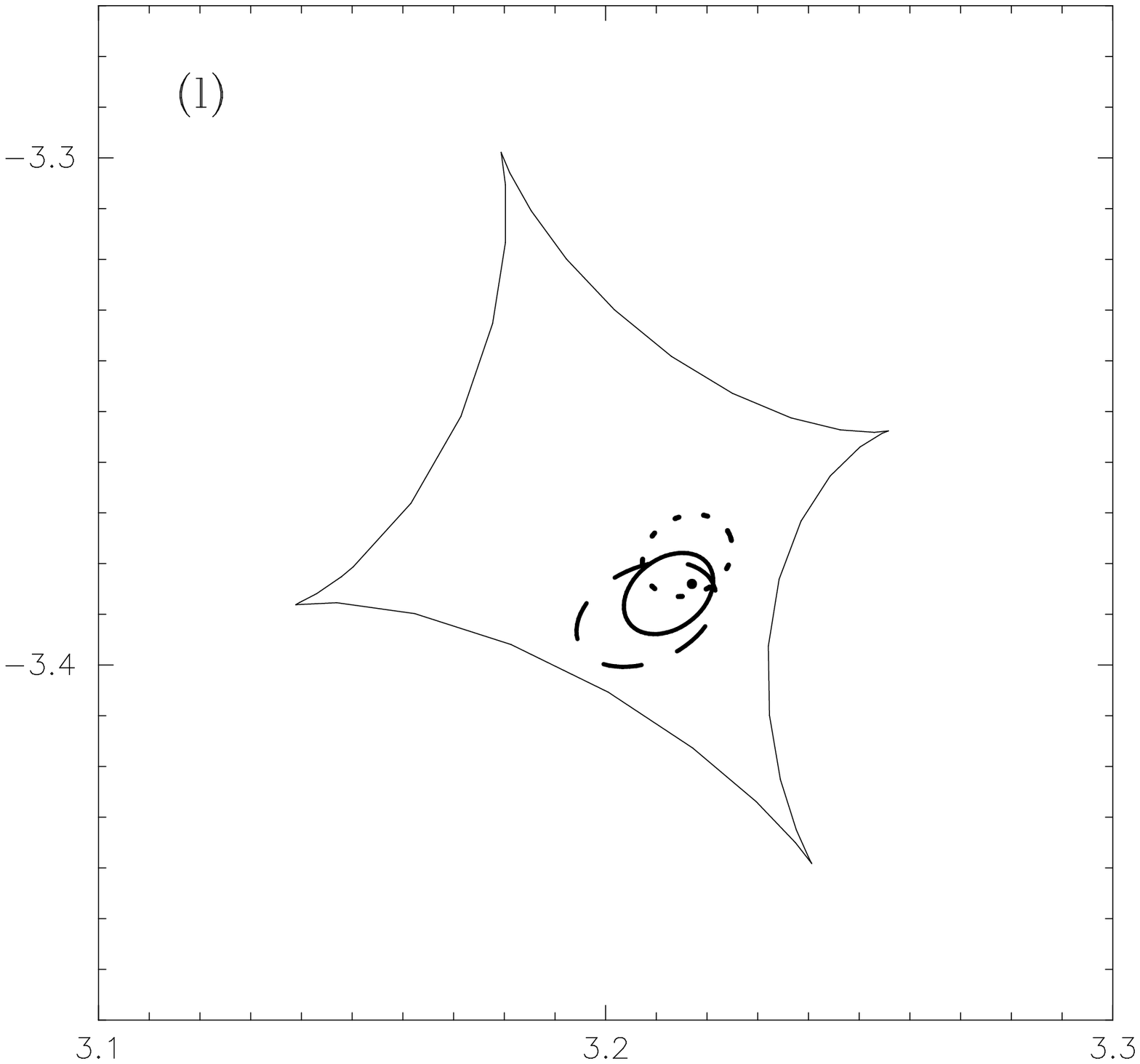},width=5.0cm}}
\end{minipage}

\caption{Results of the lens modelling of the Cloverleaf superimposed 
on the HST image. 
(a) is the
HST image overlaid with the CO observed (-225, +225 km/s);
(b) is the CO predicted for model 1, convolved by the interferometer beam.
(c) is similar as (b) but for model 2.
(j) is the the CO  predicted for model 1, not convolved by the
interferometer beam. 
(d), (e), (f) is similar as the first line but for the blue emission.
(g), (h), (i) is similar as the first line but for the red emission.
The dotted line is the corresponding critical line.
(k) gives the position of the best fitted sources for the blue (dashed),
red (dotted) and total (solid) emission, for model 1.
(l) is similar as (k) but for model 2.
The central diamond-shape curve (in (b), (e), (h), (k) and (l))
is the internal caustic crossed by the lensed CO source at redshift z=2.558.
}
\label{fig:modelco}
\end{figure*}

In order to test, to first order, the lens model with the CO(7-6) map,
we have assumed the CO source to be elliptical
with a gaussian profile. We have fitted position , size and
ellipticity so that it reproduces the observed CO image.
The upper limit of its size is provided by the CO elongated
spots A and B which are close to merging, but still clearly
separable. 
Thus, the model must predict, in the image plane,
disconnected isocontours of the A and B spots.
However, we underline that this estimated size also depends on the
ellipticity of the lens mass distribution (as the scale in the
source is proportional to $\varepsilon$R$_E$ where R$_E$ is the Einstein
radius). 
We found for model 1 a typical size of 460$\times$230 pc (FWHM) for
model 1, and 155$\times$110 pc (FWHM) for model 2.
In the present case, $1''$ in the source plane translates into
$7.68h_{50}^{-1}$ kpc with the chosen cosmology.
A summary of the CO modeling is displayed in Figure~\ref{fig:modelco}.
Figure~\ref{fig:modelco}j  shows the CO emission of 
Figure~\ref{fig:modelco}b before convolving it by the interferometer
natural beam. It clearly shows us that it will be difficult to 
reach a higher degree of precision in describing exactly the source
morphology, unless higher-resolution CO images can be acquired.
An additional effect of the convolution results in an apparent shift of the 
centroids of the CO spots with respect to the quasar point-like spots. 
This 
demonstrates that the displacements observed between the centroids of the 
CO and visible spots 
are artifacts  of the distorted morphology of the CO emission in the image
plane.

The amplication ratios inferred from the modelling and the observations
are displayed in Table~\ref{tab:modamp}. The agreement between the expected
amplifications and the observed flux ratios is relatively good.
The quasar intrinsic magnitude is about I$\sim$20.5.
The comparison between the expected amplification and the observed ones
in CO is not easy. Differential amplification due to the CO source
extent and location with respect to the diamond caustic can explain
the observed difference. The total amplification of the
CO emission is $\sim$ 18 for model 1 and $\sim$ 30 for model 2.
Following Barvainis et al (1997), these amplification factors
translate for model 1 (resp. model 2)
to molecular mass $M(H_2)$= 3. 10$^9$ M$_\odot$ (resp. 2.10$^9$
M$_\odot$) and $M(\HI)$= 3. 10$^9$ M$_\odot$ (resp. 2.10$^9$ M$_\odot$).
These mass etimates are in good agreement to the dynamical mass computed
in the next section, providing the uncertainties in the inclination and
the conversion from $M(CO)$ to $M(H_2)$.

\begin{table*}[t]
\begin{tabular}{lcccc}
\hline
Spot & Lens model 1 & Lens model 2& HST (visible) & IRAM (CO) \\
\hline
A & [2.99] 1.0 & [3.16] 1.0 & 1.0  & 1.0 \\
B & [2.84] 0.87 & [3.36] 1.20 & 0.88 $\pm$ 0.01  & 2.07 $\pm$ 0.48\\
C & [2.80] 0.84 & [2.93] 0.81 & 0.76 $\pm$ 0.01 & 1.86 $\pm$ 0.45 \\
D & [2.66] 0.74 & [2.71] 0.66 & 0.71 $\pm$ 0.01 & 1.32 $\pm$ 0.36\\
E & [-4.88] 0.0007& [-7.45] 5e-5 & - & -  \\
\hline
\end{tabular}
\caption{Amplification ratios of the four spots as inferred from the lens
modelling and compared with the visible and the CO (-225,+225 km/s)
flux ratios.
For the model we also give the $\delta$m amplification within brackets.
Accounting for the errors, the model reproduces the amplification ratios
reasonably well. The last line gives the amplification ratio of the
fifth spot. It is considerably demagnified. Since it is not detected on
the visible data, we expect its magnitude to be fainter than $m_I=25.$,
which implies that the quasar itself is fainter than $m_I=16.6$.
}
\label{tab:modamp}
\end{table*}

\subsection{Comparison with previous models}

The comparison with previous models presented by K90 and recently by Yun
et al (1997) and KKS is somewhat difficult. First, none of these 
included a lensing contribution from a distant galaxy cluster
(our model 2).
Regarding the properties of the individual lensing galaxy, the K90 
model 1 (see Table~4 and
Figure~4a of K90) and KKS are "close" to our results of model 1. 
The relative position of the
source and the lensing galaxy center are almost identical
and the orientation of the gravitational  potential is similar 
(PA of 20$^o$ for K90, 21.6$^o$ for KKS, and  21.5$^o$ for our model 1).  
The upper limit on the velocity dispersion  we find   
when no cluster component is introduced is 280 km/s, in excellent
agreement with their upper limit ($<$285 km/sec).  
But on the other hand,
our ellipticity is significantly different. K90 used the
standard definition of the eccentricity ($e=\sqrt{1-b^2/a^2}$) for the
projected mass density, the    
corresponding ellipticity is $\varepsilon=0.688$, which is 3.5 times larger 
than in our results. KKS found an ellipticity of 0.58 for an elliptical
SIS model, and  for a two shear model (elliptical SIS plus external
shear) an ellipticity of 0.40 with a external shear of 0.19.
This difference results from the analytical form of the mass profile
chosen: K90 and KKS used a SIS while we are using
a sum of two truncated EMDs with two parameters for the shape of 
the profile. 
Then, more freedom is given in the radial mass profile, leading to solutions
with smaller ellipticities (cf. the model of H14176 and discussion
in Hjorth \& Kneib 1997).

On the contrary, our results do not agree with those by Yun et al (1997)
who find a lensing elliptical potential perpendicular to 
that derived by K90 and to ours (due to the presence of a strong
external shear ?).
Indeed, the four CO spots observed (and predicted by our lens model)
better constrain the orientation of the potential (although they do not
insure the uniqueness of the solution).

\section{Constraints on the quasar CO source}

In addition to the size of the quasar CO source,
$\sim$ 460 pc FWHM, as derived above (see Sect. 5.2), we can obtain some
information on its substructure using the velocity gradient observed in
the CO(7-6) line (Figure~\ref{fig:iram}b). Indeed, the map 
(Figure~\ref{fig:iram}c) corresponding to the blue side of the line
(-225,-25 km/s) shows that spot C is stronger and slightly displaced
inwards with respect to its counterpart in the map corresponding to the
(+25,+225 km/s), shown in (Figure~\ref{fig:iram}d). A hint of this
effect can be found in Yun et al (1997) although a close comparison
of the IRAM results with the OVRO one is hampered by the fact that in
their study the spatial resolution is twice lower, the line profile
is not shown, the velocity interval considered -- 145 km/s -- is narrower 
than ours
-- 200km/s -- and not positioned precisely with respect to the line
center. 

Therefore, we shall concentrate now on the interpretation of the
IRAM data set only.
Using the two lens models obtained in Sect. 5.2, we optimized the
structure of the CO source to reproduce separately the red and the blue
images (Figure~\ref{fig:modelco}). 
We call attention to the fact that a precise registration 
(better than $0.1''$) of
the CO map and the HST data is required to achieved 
a detailed structure of the CO source. The current data does not allow
such an accurate registration, so we fine tune it so that the lens
morphology between radio and optical match is best. We tested the
registration and found that a $0.05''$ offsets does not change
the results presented below. A larger offset make the radio and optical
data inconsistent on a lens modelling point of view.

For model 1 [resp. model 2], we find an offset of 
$0.03''$ ($\sim 230$ pc) [resp. $0.02''$ ($\sim 150$ pc)]
between 
the center of regions emitting the blue and red parts of the CO line
(c.f. Figure~\ref{fig:modelco}).
We have tested this procedure against uncertainties in the CO/HST images
registration. A change by $0.05''$ (half a CO map pixel) has very minor
impact on the positions of the source regions emitting the blue and red
parts of the line. 
The quasar point-like visible source appears to be almost exactly 
centered between the blue- and red-emitting regions.
This is reminiscent of a disk- or ring-like structure
orbiting the quasar at the radius of $\sim$ 100 pc [resp. 75 pc]
and with a Keplerian
velocity of $\sim$ 100 km/s (assuming a 90 deg inclination with respect to
the plane of the sky), the resultating central mass would be
$\sim$ 10$^{9}$ M$_\odot$ [resp. $\sim$ 7.5 10$^{8}$ M$_\odot$].
Elaborating a more sophisticated (realistic) model of the CO source and
its link  with the BAL feature also observed in this quasar is beyond
the current analysis but should be performed in the future.
Regarding the spot fluxes in CO(7-6) as derived from 
the maps in Figures~\ref{fig:iram}a,
\ref{fig:iram}c, \ref{fig:iram}d and Table~\ref{tab:spotiram}, a
comparison with the fluxes given previously in Alloin et al (1997)
is not straightforward because the velocity interval over which the
integration has been performed is different in the two studies
(-225,+225 km/s vs. -325,325 km/s). The amplification factor being
extremely sensitive to the position of the emitting region with respect
to the caustic, any velocity-positional changes within the source can
result in a different configuration in the image plane. Such gradients
are likely to occur in the molecular torus of the standard AGN model.
The differential amplification resulting from this gradient is probably
at the origin of the asymmetry observed in the line profile 
(Figure~\ref{fig:lineprofile}).

\section{Discussion, Prospective and Conclusion} 

The complete multi-wavelength analysis of the Cloverleaf reveals that 
this is probably a complex lens which includes a lensing galaxy and an
additional distant lensing cluster of galaxies.  
The reality of the cluster toward the Cloverleaf has still to be
confirmed independently. 
Even at the level of a 4 $\sigma$
detection, clustering and projection effects can not be 
readily discarded to explain  the observed galaxy enhancement, and
measuring the redshift of these galaxies is of high priority to
position them in redshift space.

Yet, most of the galaxies around the
Cloverleaf are found in the same magnitude and size ranges, as expected 
if they indeed belonged to a cluster.  If the cluster is at a very
large distance, the shift of its galaxy      
luminosity function up to higher apparent magnitude would  explain 
why the number-density contrast of the cluster with respect to faint
field galaxies is lowered down to only a 4 $\sigma$ level.

This interpetation implies that the lensing galaxy may not be very massive 
and consequently may not be very luminous, helping to explain the mystery 
of the lensing galaxy not having been detected so far.   The
drawback is that, despite the constraint that the shapes of the 
CO spots provide on the orientation of the mass density distribution,  
it mandates sharing the 
mass between the lensing-galaxy and the lensing-cluster which  
 increases the number of possible lens configurations.
 Hence, it considerably reduces the chances to infer 
a secure measure of the Hubble constant from
the time delay measurements of lightcurves between the four
spots (a thorough report of the variability of the quasar is given
in \O stensen et al. 1997). A measure of time-delay would bring
additional constraint on the mass distributioin of this system.
Thus, we need to confirm, probably from ultra-deep visible 
images and spectra and additional near infrared photometry, that the cluster is
there and that it is at a large redshift.  

Ultra-deep visible and K band
images might also reveal the position and the shape of the light
distribution of the lensing galaxy. Such information
would be useful to improve the mapping of the CO
source, as already emphasised by Yun et al. (1997)
and Alloin et al. (1997). With the present-day data, 
the CO source is found to be a disk- or ring-like structure with typical radius
of $\sim$ 100 pc, under the lens model of a galaxy and a cluster at
z=1.7, leading to a central $\sim$ 10$^{9}$ M$_\odot$ object, typical 
of a black-hole.
It is amazing to see that a disk with such a
small intrinsic size can be spatially 
``resolved" even at an angular distance as large as 1.6 $h^{-1}_{50}$ Gpc.
However, we are aware that the CO source could have a more complex geometry 
and as long as the exact redshift and mass distribution of the
lenses will not be known in a direct way, uncertainties on the CO source 
size and structure will persist.

Although this remains to be confirmed independently, the discovery of a 
distant cluster of galaxies on the line of sight to
the Cloverleaf is remarkable because it reinforces the suspicion that 
many bright 
high redshift quasars are magnified by cluster-like systems at large 
distances. This was already reported from analyses in the fields of the 
doubly imaged quasar Q2345+007 (Bonnet et al
1993; Mellier et al 1994; van Waerbeke et al 1997), 
where a cluster candidate is expected to be at z$\sim$ 0.75 
(Pell\'o et al 1996) and of MG2016 where the X-ray emission 
of the intra-cluster gas has been observed and for which the Iron 
line (from X-ray spectroscopy) gives a redshift z$\sim$1 
(Hattori et al 1997).\\

These cases of strong lensing may substantially change the intrinsic bright-end
luminosity function of quasars. In fact, there is now convincing
evidence that magnification biases play an important role and this should draw some important 
cosmological issues. Early observational 
evidence was emphazised by  the
galaxy-quasar associations detected by Fugmann (1990). They 
have been re-investigated and confirmed by Bartelmann
\& Schneider (1994) and Be\~nitez \& Mart\'{\i}nez-Gonzalez  (1996). 
 According to Bartelmann \& 
Schneider (1992) the galaxy-quasar association on arcminute scale 
cannot be explained by lensing effects from single galaxies only, but 
must involve lensing effects by 
large-scale structure or  rich clusters of galaxies.  Since all these sources,
including the Cloverleaf, are very bright, they correspond to
preferentially selected fields which probe the bright-end of the
magnified areas where we expect that the lensing agent responsible for
magnification must be strong.  If so, many bright quasars 
are magnified by massive gravitational systems.  It would therefore be 
important to collect ultra-deep visible and near infrared high-resolution
imaging on a large sample of overbright quasars in order to
compute what fraction is surrounded by high-redshift foreground 
clusters of galaxies.

Though these fields correspond to biased lines of
sight, they are typical fields showing the  strong-end of cosmic shear 
events expected on arcminute scales from the predictions of Jain \& Seljak 
(1996).  By using the non-linear evolution of the power spectrum,
these authors have shown that the rms cosmic shear on such scales
is more than twice the values predicted from the  weakly 
non-linear regime. Arguing that the cosmic shear should therefore be 
observable even with present-day ground
based telescopes, Schneider et al (1997) have re-analysed the Fort et al (1996) data 
and have shown that the cosmic shear may have already been detected in at
least one quasar field (PKS1508). Very 
deep observations of the Cloverleaf in the near IR, to confirm that a 
shear pattern with significant shear amplitude of order of 5\% is present
around the four spots, would provide independent data in favor of the   
this interpretation.  

The observation of cosmic shear or the detection 
of distant clusters puts strong constraints on the density parameter $\Omega$ 
and on the initial power spectrum of density fluctuations. If most of the 
bright quasars are magnified by high-redshift
clusters, then the density of distant clusters of galaxies must be large
(this may conflict with the standard CDM model for example).  
However, the impact on cosmological scenarios is not clear since we do not 
understand yet the selection biases.  

Though our multiwavelength analysis, using the best visible and CO images,
provides a new interpretation of the Cloverleaf and a good understanding
of the lens configuration, the discussion raises new questions
about the CO source and the lenses. 
In order to address these questions thouroughly, additional photometric
and spectroscopic data are required.
Since the members of the cluster
are extremely faint their spectroscopy 
and the measurement of the cluster
velocity dispersion will be technically difficult. In addition to
ultra-deep visible and near infrared imaging to probe the light 
distribution of the lenses (individual galaxy and cluster),   
 the mass model representing the cluster could be considerably improved 
by using the shear pattern generated by the lensed background sources and 
from the redshift measurements of the `brightest' cluster members. 
Last but not least, the photometric monitoring of the four spots 
(\O stensen et al 1997) will provide the time delay of the light curves 
which is also an important and independent constraint for the mass modelling.

\acknowledgements{We thank  M.-C. Angonin and C. Vanderriest 
for providing their CFHT images. We thank J. B\'ezecourt, S. Charlot,  
J.-M. Miralles and  R. Pell\'o, for usefull discussions and their 
help in using and interpreting properly the Bruzual \& Charlot'
galaxy evolution models.  
We thank A. Lannes for useful discussion on deconvolution and inversion
techniques, and J. Hjorth on multiple quasar lensing and PSF substraction.
DA and YM thank Observatoire Midi-Pyr\'en\'ees for hospitality.
We are grateful to  CEA,  CFHT, IRAM, ST-ECF, CADC and
Cambridge (APM)  for financial and technical supports. 
}

\end{document}